\documentclass[12pt]{article}

\usepackage{amsmath}
\usepackage{amsfonts}
\usepackage{graphicx}
\usepackage[a4paper, total={18cm, 23cm}]{geometry}
\usepackage{multirow}
\usepackage{tikz}
\usetikzlibrary{arrows}
\usetikzlibrary{decorations.markings}
\usepackage{color}
\usepackage{soul}
\usepackage{colortbl}
\usepackage{rotating}
\usepackage{natbib}

\begin{document}

\title{\textbf{\Large{Relationship between Collider Bias and Interactions on the Log-Additive Scale}}}
\author{Apostolos Gkatzionis$^{1}$, Shaun R. Seaman$^{3}$, Rachael A. Hughes$^{1,2}$, Kate Tilling$^{1,2}$}
\date{\begin{small}$^1$MRC Integrative Epidemiology Unit, University of Bristol, UK. \\
$^2$Population Health Science Institute, Bristol Medical School, University of Bristol, UK. \\
$^3$MRC Biostatistics Unit, University of Cambridge, Cambridge, UK.
\end{small}}
\maketitle

\abstract{Collider bias occurs when conditioning on a common effect (collider) of two variables $X, Y$. In this manuscript, we quantify the collider bias in the estimated association between exposure $X$ and outcome $Y$ induced by selecting on one value of a binary collider $S$ of the exposure and the outcome. In the case of logistic regression, it is known that the magnitude of the collider bias in the exposure-outcome regression coefficient is proportional to the strength of interaction $\delta_3$ between $X$ and $Y$ in a log-additive model for the collider: $\mathbb{P} (S = 1 | X, Y) = \exp \left\{ \delta_0 + \delta_1 X + \delta_2 Y + \delta_3 X Y \right\}$. We show that this result also holds under a linear or Poisson regression model for the exposure-outcome association. We then illustrate by simulation that even if a log-additive model with interactions is not the true model for the collider, the interaction term in such a model is still informative about the magnitude of collider bias. Finally, we discuss the implications of these findings for methods that attempt to adjust for collider bias, such as inverse probability weighting which is often implemented without including interactions between variables in the weighting model.}


\section{Introduction}

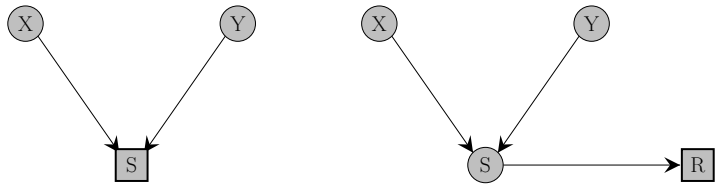
\begin{figure}[h]
\begin{center}
\begin{tikzpicture}[scale=0.5, every node/.style={scale=0.5}]
\draw [decoration={markings,mark=at position 1 with
    {\arrow[scale=2,>=stealth]{>}}},postaction={decorate}] (-7.65, 1.65) -- (-5.35, -1.65);
\draw [decoration={markings,mark=at position 1 with
    {\arrow[scale=2,>=stealth]{>}}},postaction={decorate}] (-2.35, 1.65) -- (-4.65, -1.65);
\draw[thick, fill = lightgray] (-5.45, -1.55) rectangle (-4.55, -2.45);
\draw[fill = lightgray] (-8, 2) circle (0.5cm);
\draw[fill = lightgray] (-2, 2) circle (0.5cm);
\node (a) at (-5, -2) {\Large{S}};
\node (b) at (-8, 2) {\Large{X}};
\node (c) at (-2, 2) {\Large{Y}};

\draw [decoration={markings,mark=at position 1 with
    {\arrow[scale=2,>=stealth]{>}}},postaction={decorate}] (2.35, 1.65) -- (4.65, -1.65);
\draw [decoration={markings,mark=at position 1 with
    {\arrow[scale=2,>=stealth]{>}}},postaction={decorate}] (7.65, 1.65) -- (5.35, -1.65);
\draw [decoration={markings,mark=at position 1 with
    {\arrow[scale=2,>=stealth]{>}}},postaction={decorate}] (5.5, -2) -- (10.5, -2);
\draw[thick, fill = lightgray] (10.55, -1.55) rectangle (11.45, -2.45);
\draw[fill = lightgray] (2, 2) circle (0.5cm);
\draw[fill = lightgray] (8, 2) circle (0.5cm);
\draw[fill = lightgray] (5, -2) circle (0.5cm);
\node (d) at (5, -2) {\Large{S}};
\node (e) at (2, 2) {\Large{X}};
\node (f) at (8, 2) {\Large{Y}};
\node (g) at (11, -2) {\Large{R}};
\end{tikzpicture}
\end{center}
\caption{Causal diagram illustrating how collider bias occurs. Two variables $X$, $Y$ that are both causes of a third variable $S$ will become correlated when conditioning on $S$ (left) or on a variable that is causally downstream of $S$ (right) even if they were unconditionally independent. Or, if $X$ and $Y$ are unconditionally associated, conditioning on $S$ will change the strength of their association.} 
\label{dag1}
\end{figure}

Collider bias is a common concern in epidemiological studies. When exploring the association between an exposure $X$ and an outcome $Y$ of interest, collider bias occurs if the analysis is conditioned on a common effect of the exposure and outcome, or a variable that is causally downstream of the common effect (a ``child" of the collider), as illustrated in Figure~\ref{dag1}. Numerous examples of studies affected by collider bias can be found in the literature. For example, collider bias has been suggested as an explanation for the ``obesity paradox", where obesity often appears to be associated with decreased mortality in older individuals or people suffering from chronic diseases, despite being associated with increased mortality in the overall population \citep{Sperrin2016, Viallon2016}.

In this paper, we focus on collider bias induced when an analysis is restricted to a single level of a binary collider variable $S$. A number of biases in epidemiological studies across a wide range of study designs can be attributed to this mechanism. This includes some forms of selection bias due to non-representative sampling \citep{Hernan2004}, survival bias, Berkson's hospitalization bias \citep{Berkson1946} and index event bias in studies of disease progression \citep{Mitchell2022}, among others. In all of these cases, the collider $S$ represents selection into the study: individuals with $S = 1$ have their exposure and outcome observed, while individuals with $S = 0$ do not and are hence excluded from the study.

It has long been recognized in the literature that this type of collider bias relates to interactions between the exposure and the outcome on the log-additive scale in their effects on the collider \citep{Greenland1977, Kleinbaum1982}. Consider the following log-additive model for the collider:
\begin{equation}
	\log \mathbb{P} (S = 1 | X, Y) = \delta_0 + \delta_1 X + \delta_2 Y + \delta_3 X Y \label{logadd}
\end{equation}
The parameter $\delta_3$ quantifies the strength of the exposure-outcome interaction. In analyses of binary outcomes, a number of authors have suggested that collider bias will not affect estimates of the exposure-outcome odds ratio when $X$ and $Y$ do not interact in their effects on $S$, i.e. when $\delta_3 = 0$ \citep{Greenland1996, Rothman2008, Greenland2009, VanderWeele2015, Bartlett2015}. An overview of the relevant literature can be found in \cite{Jiang2017}, who also explored the direction in which collider bias acts, while \cite{Mansournia2022} obtained similar results and also investigated bias when conditioning on $S = 0$. Despite the interest this topic has attracted, most of the relevant papers have focused on binary outcome variables, with some papers also restricting the exposure to be a binary variable. One exception is \cite{Shahar2017}, who considered a discrete exposure and a discrete outcome, potentially with more than two categories. 

In this manuscript, we expand the literature by considering a wider range of outcome variables, including count outcomes (Poisson regression) and continuous outcomes (linear regression) in addition to binary ones. We also place no conditions on the form of the exposure variable. We show that collider bias will not affect the exposure-outcome association when the exposure and outcome do not interact in their effects on the collider, i.e. when $\delta_3 = 0$ in model \eqref{logadd}. When they do interact, we show that the magnitude of collider bias induced in the exposure-outcome regression coefficients in linear, logistic and Poisson regression models is proportional to $\delta_3$. Finally, when model \eqref{logadd} is misspecified, we show numerically that for binary $X$ there is still a linear relationship between the magnitude of the collider bias and the estimated value of $\delta_3$ in the (misspecified) model \eqref{logadd}.

The rest of this manuscript is organized as follows. In Section 2, we review the relevant literature. We then consider regression models for a binary outcome $Y$ (primarily logistic regression), linear regression and Poisson regression. For each of these models, we prove that the magnitude of collider bias is proportional to the value of the exposure-outcome interaction $\delta_3$ in the log-additive model \eqref{logadd}. In Section 3, we investigate collider bias when the collider variable is not distributed according to the log-additive model \eqref{logadd}. We demonstrate numerically that when $X$ is a binary variable, estimating $\delta_3$ can still provide information about the magnitude of collider bias, even if model \eqref{logadd} is not the true model for $S$. Section 4 contains an illustrative application using data from the Avon Longitudinal Study of Parents and Children (ALSPAC) to investigate associations between maternal traits, such as education or smoking before pregnancy, and child sex. A discussion of our main findings and their implications is presented in Section 5.

\section{Collider Bias under a Log-additive Model for the Collider}

\subsection{Statement of the Problem}

As in the introduction, suppose that the objective is to investigate the association between an exposure $X$ and an outcome $Y$. We focus on the marginal association between $X$ and $Y$.  However, the results presented here can be extended to cover the conditional association between $X$ and $Y$ given a set of other variables, e.g. confounders, as we briefly discuss towards the end of Section 2.2.1.

As in the previous section, we let the collider $S$ represent a binary selection indicator. The unconditional exposure-outcome association cannot be estimated directly using the observed data; instead, only the conditional association given $S = 1$ can be estimated. As stated in Section 1, if an individual's exposure and outcome values affect their likelihood of selection into the study sample, the conditional and unconditional exposure-outcome associations will differ. Our aim in this section is to explore the difference between the conditional and unconditional exposure-outcome associations for a binary collider variable distributed according to model \eqref{logadd}. 

We study collider bias separately for binary, continuous and count outcome variables. For binary outcome variables, we examine collider bias in odds ratios, logistic regression coefficients and risk ratios. For continuous outcomes, we quantify collider bias in linear regression coefficients, and for count outcome variables, we investigate collider bias in the log rate ratio parameters of a Poisson regression model. We work under the assumption that the collider $S$ is distributed according to the log-additive model \eqref{logadd}; different models for $S$ will be considered in the next section.

\subsection{The Relationship between Collider Bias and Exposure-Outcome Interactions}

\subsubsection{Binary Outcome - Collider Bias on the Odds Ratio Scale}

Consider first the case of a binary outcome, which has received the most attention in the literature \citep[e.g.][]{Jiang2017, Bartlett2015}. Some of these papers also restricted the exposure to be binary; here, we do not place any assumptions on the type of the exposure variable. Our main assumption is that the exposure and outcome affect the collider on the log-additive scale, as in \eqref{logadd}. Let
\begin{eqnarray}
	OR_{XY} (x) & = & \frac{\mathbb{P} (Y = 1 | X = x + 1)}{\mathbb{P} (Y = 0 | X = x + 1)} \times \frac{\mathbb{P} (Y = 0 | X = x)}{\mathbb{P} (Y = 1 | X = x)} \nonumber \\
	OR_{XY | S = 1} (x) & = & \frac{\mathbb{P} (Y = 1 | X = x + 1, S = 1)}{\mathbb{P} (Y = 0 | X = x + 1, S = 1)} \times \frac{\mathbb{P} (Y = 0 | X = x, S = 1)}{\mathbb{P} (Y = 1 | X = x, S = 1)} \nonumber
\end{eqnarray}
be the unconditional and conditional odds ratios respectively.

In Supplementary Section 1.1 \citep[see also][]{Jiang2017}, we prove that
\begin{equation}
	 OR_{XY | S = 1} (x) = OR_{XY} (x) \; \exp \left\{ \delta_3 \right\} \label{bias.or}
\end{equation}
This shows that the magnitude of collider bias on the odds ratio scale is fully determined by the interaction parameter $\delta_3$. In particular, if $\delta_3 = 0$, we have $OR_{XY | S = 1} (x) = OR_{XY} (x)$, meaning that collider bias does not occur. It follows from equation \eqref{bias.or} that if the logistic model
\begin{equation}
	\text{logit} \mathbb{P} (Y = 1 | X) = \beta_0 + \beta_1 X \label{logistic.model}
\end{equation}
is correctly specified, the magnitude of collider bias in the parameter $\beta_1$ is equal to $\delta_3$. Letting $\beta_1^S$ denote the (population) log-odds ratio for the exposure-outcome association conditional on $S = 1$, one could write
\begin{equation}
	\beta_1^S = \beta_1 + \delta_3 \label{bias.logistic}
\end{equation}

Some generalizations of this result are possible. For example, equation \eqref{bias.logistic} still holds when the $X-S$ and $Y-S$ main effects in model \eqref{logadd} are replaced by non-linear functions:
\begin{equation*}
	\log \mathbb{P} (S = 1 | X, Y) = g_1 (X) + g_2 (Y) + \delta_3 X Y
\end{equation*}
Finally, consider the more general case of a higher-order (in X) interaction:
\begin{equation*}
	\log \mathbb{P} (S = 1 | X, Y) = g_1 (X) + g_2 (Y) + g_3 (X) Y
\end{equation*}
In this case, equation~\eqref{bias.or} becomes
\begin{equation*}
	OR_{XY | S = 1} (x) = OR_{XY} (x) \exp \left\{ g_3 (x + 1) - g_3 (x) \right\}
\end{equation*}
The pattern of bias will therefore depend on the form of the function $g_3$. Note that the conditional and unconditional odds ratios will be equal if and only if $g_3 (x + 1) = g_3 (x)$ for all $x$, i.e. $g_3$ is constant in $X$, which again shows that collider bias occurs if and only if the exposure and outcome interact in their effects on the collider.

Finally, let the variable $X$ be vector-valued; this may represent either multiple exposures whose association with $Y$ is investigated, or a single exposure whose association with the outcome is adjusted for the presence of observed confounders. Under the models \eqref{logadd} and \eqref{logistic.model}, where now $\beta_1, \delta_1, \delta_3$ are vector-valued, one can show that $\beta_1^S = \beta_1 + \delta_3$, i.e. the bias in the regression coefficient of an element $X_j$ of the vector $X$ is equal to the interaction between $X_j$ and $Y$ in the collider model. Moreover, interactions between the variables $X_j$ in the collider model will not affect the bias.

\subsubsection{Binary Outcome - Collider Bias on the Risk Ratio Scale}

We now explore the magnitude of collider bias on the risk ratio scale. Once again, we assume that the collider $S$ is distributed according to \eqref{logadd} and that the outcome $Y$ is binary. Our aim is to compare the unconditional risk ratio
\begin{equation*}
	RR_{XY} (x) = \frac{\mathbb{P} (Y = 1 | X = x + 1)}{\mathbb{P} (Y = 1 | X = x)}
\end{equation*}
to the conditional risk ratio
\begin{equation*}
	RR_{XY | S = 1} (x) = \frac{\mathbb{P} (Y = 1 | X = x + 1, S = 1)}{\mathbb{P} (Y = 1 | X = x, S = 1)}
\end{equation*}
In Supplementary Section 1.2, we prove that
\begin{equation}
	RR_{XY | S = 1} (x) = RR_{XY} (x) \times \frac{ e^{\delta_2 + \delta_3 (x + 1)} \; \mathbb{P} (Y = 1 | X = x) + e^{\delta_3} \; \mathbb{P} (Y = 0 | X = x) }{ e^{\delta_2 + \delta_3 (x + 1)} \; \mathbb{P} (Y = 1 | X = x + 1) + \mathbb{P} (Y = 0 | X = x + 1) } \label{bias.rr}
\end{equation}
It is clear from this formulation that the properties we have proved for the magnitude of collider bias on the odds ratio scale do not hold for bias on the risk ratio scale; for example, unlike \eqref{bias.logistic}, expression \eqref{bias.rr} involves both the interaction parameter $\delta_3$ and the outcome-collider coefficient $\delta_2$.  A more specific formula for the bias on the risk ratio scale can be obtained by incorporating modelling assumptions for the exposure-outcome relationship into \eqref{bias.rr}. In practice, risk ratios are often studied under the log-binomial regression model
\begin{equation*}
	\text{log} \mathbb{P} (Y = 1 | X = x) = \beta_0 + \beta_1 x
\end{equation*}
This implies an unconditional risk ratio of $RR_{XY} (x) = e^{\beta_1}$ for any $x$, and from \eqref{bias.rr}, a conditional risk ratio of
\begin{equation*}
	RR_{XY | S = 1} (x) = \frac{ e^{\delta_2 + \delta_3 (x + 1)} \; e^{\beta_0 + \beta_1 (x + 1)} + e^{\delta_3 + \beta_1} \; (1 - e^{\beta_0 + \beta_1 x}) }{ e^{\delta_2 + \delta_3 (x + 1)} \; e^{\beta_0 + \beta_1 (x + 1)} + (1 - e^{\beta_0 + \beta_1 (x + 1)}) }
\end{equation*}
The difference between the unconditional and conditional risk ratios is then
\begin{equation}
	RR_{XY} (x) - RR_{XY | S = 1} (x) = e^{\beta_1} \; \left( 1 - \frac{ e^{\delta_2 + \delta_3 (x + 1)} \; e^{\beta_0 + \beta_1 x} + e^{\delta_3} \; (1 - e^{\beta_0 + \beta_1 x}) }{ e^{\delta_2 + \delta_3 (x + 1)} \; e^{\beta_0 + \beta_1 (x + 1)} + (1 - e^{\beta_0 + \beta_1 (x + 1)}) } \right) \label{bias.logit}
\end{equation} 
The bias hence depends on both the interaction term $\delta_3$ and the outcome-collider parameter $\delta_2$, and the absence of an interaction ($\delta_3 = 0$) is not enough to eliminate bias on the risk ratio scale. On the other hand, one can easily verify that the conditional and unconditional risk ratios are equal when $\delta_2 = \delta_3 = 0$, i.e. when the outcome does not associate with the collider; and likewise, there is no bias if $\delta_3 = 0$ and $\beta_1 = 0$.

In Supplementary Section 1.2, we explore collider bias on the risk ratio scale under a logistic regression model for the outcome and obtain results similar to those reported here.

\subsubsection{Continuous Outcome - Collider bias in Linear Regression Coefficients}

We now turn our attention to continuous outcome variables, and assume that the outcome is distributed according to the linear regression model
\begin{equation*}
	Y | X = \beta_0 + \beta_1 X + \epsilon_Y
\end{equation*}
where  $\epsilon_Y \sim N(0, \sigma^2)$ independent of $X$. We start by noting that $\mathbb{E} (Y | X) = \beta_0 + \beta_1 X$ and explore the bias in the regression coefficient $\beta_1$ when conditioning on $S = 1$. In Supplementary Section 1.3, we show that the conditional expectation $\mathbb{E} (Y | X, S = 1)$ is equal to
\begin{equation}
	\mathbb{E} (Y | X, S = 1) = (\beta_0 + \delta_2 \sigma^2) + (\beta_1 + \delta_3 \sigma^2) x
\end{equation}
Denoting by $\beta_0^S$ and $\beta_1^S$ the regression coefficients of a linear regression model conditioned on $S = 1$, it follows that
\begin{equation}
	\beta_1^S = \beta_1 + \delta_3 \sigma^2 \label{bias.linear}
\end{equation}
which implies that the two regression coefficients in the conditional and unconditional exposure-outcome models will be equal if and only if $\delta_3 = 0$. When the two coefficients differ, the magnitude of collider bias induced is equal to the interaction term $\delta_3$ multiplied by the outcome variance $\sigma^2$. Moreover, $\beta_0^S = \beta_0 + \delta_2 \sigma^2$; therefore the bias in the intercept $\beta_0$ is equal to the outcome-collider parameter $\delta_2$ multiplied by the outcome variance.

As with binary outcomes, the above derivation allows for a non-linear exposure-collider effect,
\begin{equation*}
	\log \mathbb{P} (S = 1 | X, Y) = g_1 (X) + \delta_2 Y + \delta_3 X Y
\end{equation*}
but not for a non-linear outcome-collider effect. Finally, a more general regression framework for the exposure-outcome relationship can be considered:
\begin{equation*}
	Y | X = x = m (x; \beta) + \epsilon \;\;\; , \;\;\; \epsilon \sim N(0, \sigma^2)
\end{equation*}
where $m(x; \beta) = \mathbb{E} (Y | X = x)$ is a potentially non-linear function that represents the exposure-outcome association. This yields
\begin{equation*}
		\mathbb{E} (Y | X = x, S = 1) = m (x; \beta) + \sigma^2 \delta_2 + \sigma^2 \delta_3 x
\end{equation*}
In addition, as for logistic regression, our results still hold if $X$ is vector-valued.

\subsubsection{Count Outcome - Collider bias in Poisson Regression Coefficients}

Finally, we consider the case of a count outcome variable distributed according to the Poisson regression model
\begin{equation}
	Y | X \sim Poisson (\lambda) \;\;\; , \;\;\; \lambda = \lambda (X) = \exp \left\{ \beta_0 + \beta_1 X \right\} \label{poisson.model}
\end{equation}
Once again, our aim is to obtain an expression for the bias in the regression coefficient $\beta_1$ when the collider $S$ follows the log-additive model \eqref{logadd}. Our framework here has some similarities to the work of \cite{Shahar2017}; their paper requires that the exposure is a discrete variable but does not place any distributional assumptions on the outcome, apart from it being discrete.

In Supplementary Section 1.4, we show that if expression \eqref{poisson.model} holds, then $Y | X, S = 1$ is Poisson($\kappa$)-distributed, where
\begin{equation}
	\kappa = \kappa (X) = \exp \left\{ (\beta_0 + \delta_2) + (\beta_1 + \delta_3) X \right\} \label{poisson.cond}
\end{equation}
Therefore, the relationship between the regression coefficients $\beta_0$ and $\beta_1$ in the unconditional exposure-outcome model and the corresponding coefficients $\beta_0^S$ and $\beta_1^S$ in the conditional model is
\begin{eqnarray}
	\beta_0^S & = & \beta_0 + \delta_2 \nonumber \\
	\beta_1^S & = & \beta_1 + \delta_3 \label{bias.poisson}
\end{eqnarray}
As in the case of logistic regression \eqref{bias.logistic}, this implies that the magnitude of collider bias in the regression coefficient $\beta_1$ induced by conditioning on $S = 1$ is equal to $\delta_3$. In particular, when the exposure and outcome do not interact in their effects on the collider in the log-additive model \eqref{logadd}, i.e. $\delta_3 = 0$, there is no bias.

As with binary and continuous outcome variables, a few extensions of this result are possible, including to analyses with a non-linear exposure-collider association: if expression \eqref{poisson.model} holds and
\begin{equation*}
	\log \mathbb{P} (S = 1 | X, Y) = g_1 (X) + \delta_2 Y + g_3 (X) Y
\end{equation*}
then $Y | X, S = 1 \sim Poisson (\kappa (x))$ where $\kappa (x) = \exp \{ \beta_0 + \beta_1 x + \delta_2 + g_3 (x) \}$.
Finally, our results can be readily extended to Poisson regression with a vector-valued exposure variable.

\section{Collider Bias under Alternative Models for the Collider - An Asymptotic Study}

\subsection{Study Design}

So far we have assumed that the collider $S$ is distributed according to the log-additive model \eqref{logadd}. Under this assumption, we have shown that there is a linear relationship between the magnitude of collider bias and the interaction term $\delta_3$ in model \eqref{logadd}, as seen in equations \eqref{bias.logistic}, \eqref{bias.linear} and \eqref{bias.poisson}. However, the log-additive model \eqref{logadd} may be misspecified. In this section, we conduct an asymptotic study to investigate collider bias under misspecification of model \eqref{logadd}. Focusing on the case of a binary exposure variable, we demonstrate two things.  First, if the true model for $S$ is not log-additive, the exposure-outcome interaction term in that model does not exhibit a linear relationship with the magnitude of collider bias in exposure-outcome regression coefficients. Second, the limiting value (as the sample size tends to infinity) of the maximum likelihood estimator (MLE) of the interaction parameter $\delta_3$ obtained by fitting the log-additive model \eqref{logadd} for $S$ still exhibits a linear relationship with the magnitude of collider bias, even if this log-additive model is misspecified. 

For our asymptotic study, we considered nine data generating mechanisms, obtained by combining three different outcome models and three different models for the selection indicator $S$. For the outcome, we used linear, logistic and Poisson regression:
\begin{eqnarray}
	(Y1) & : & \text{logit} \mathbb{P} (Y = 1 | X) = \beta_0^{Y1} + \beta_1^{Y1} X \nonumber \\
	(Y2) & : & Y | X = x = \beta_0^{Y2} + \beta_1^{Y2} x + \epsilon_Y \;\;\; , \;\;\; \epsilon_Y \sim N(0, \sigma^2) \nonumber \\
	(Y3) & : & Y | X = x \sim Poisson (\lambda) \;\;\; , \;\;\; \lambda = \lambda (x) = \exp \left\{ \beta_0^{Y3} + \beta_1^{Y3} x \right\} \nonumber
\end{eqnarray}
For the selection indicator $S$, we considered standard logistic and probit regression models:
\begin{eqnarray}
	(S1) & : & \mathbb{P} (S = 1 | X, Y) \; = \; \text{expit} \left\{ \delta_0^{S1} + \delta_1^{S1} X + \delta_2^{S1} Y + \delta_3^{S1} X Y \right\} \nonumber \\
	(S2) & : & S \; = \; \mathbf{1}_{S' > 0} \;\;\; , \;\;\; S' \sim N \left( \delta_0^{S2} + \delta_1^{S2} X + \delta_2^{S2} Y + \delta_3^{S2} X Y, 1.6^2 \right) \nonumber
\end{eqnarray}
We also considered a third model, where we generated a latent, normally distributed variable $S'$ and then set $S = 1$ for individuals for which the latent variable took values below a lower threshold $r_1$ or above an upper threshold $r_2$:
\begin{equation*}
	(S3) \;\;\; : \;\;\; S = \mathbf{1}_{S' > r_1 \; \text{or} \; S' < r_2} \;\;\; , \;\;\; S' \sim N \left( \delta_0^{S3} + \delta_1^{S3} X + \delta_2^{S3} Y + \delta_3^{S3} XY, 1.6^2 \right)
\end{equation*}
We will refer to (S3) as a ``double-threshold" model. Finally, in all nine data generating mechanisms, the exposure values were generated from a $\text{Bernoulli} (0.3)$ distribution.

The parameters of the three outcome models and the three data-generating models for $S$ were specified as follows. In the outcome model, we set $\beta_0^{Yj} = 0$ and $\beta_1^{Yj} = 0.2$, $j = 1, 2, 3$. In the linear regression model (Y2), we also set $\sigma = 0.5$. In the model for $S$, the exposure-collider and outcome-collider association parameters were set to $\delta_1^{Sk} = \delta_2^{Sk} = 0.3$, $k = 1, 2, 3$. Finally, the residual variance for the latent variable $S'$ in models (S2) and (S3) was set to $1.6$ so that the regression coefficients in these models represented a comparable strength of association as the coefficients of the logistic regression model (S1) \citep[][, Chapter 17]{Wooldridge2012}.

There were two parameters to be varied in our asymptotic study: the strength of the exposure-outcome interaction $\delta_3^{Sk}$, and the proportion of selected individuals, which was determined by the intercept term $\delta_0^{Sk}$ in the model for $S$ (and for model (S3), by the thresholds $r_1, r_2$). We conducted two experiments varying the values of these parameters. In the first experiment, we specified the intercept term $\delta_0^{Sk}$ so that approximately $50\%$ of simulated individuals were included in the conditional analysis ($S = 1$); in model (S3), we instead set $\delta_0^{S3} = 0$ and set $r_1, r_2$ equal to the first and third quartile of the distribution of $S'$. We then varied the exposure-outcome interaction parameter $\delta_3^{Sk}$, letting it take the values $\delta_3^{Sk} = 0, \pm 0.1, \pm 0.2, \pm 0.3, \pm 0.4, \pm 0.5$. In our second experiment, we used the same range of values for $\delta_3^{Sk}$ but varied the value of the intercept $\delta_0^{Sk}$. We specified four values for $\delta_0^{Sk}$, such that the proportion of selected individuals was $10\%$, $30\%$, $70\%$ and $90\%$ respectively. For model (S3), we set $\delta_0^{S3} = 0$ and specified the proportion of selected individuals by tuning the thresholds $r_1, r_2$ instead.

For each of the nine data generating mechanisms and each set of parameter values, we generated a single dataset of size $n = 10^7$. We used a large sample size to approximate an infinite sample: the values of maximum likelihood estimators obtained using our sample will be very close to the limiting value of these estimators. We then fitted model (Yj) using only data on individuals with $S = 1$ to calculate the estimate $\hat{\beta}_1^{Yj}$ of $\beta_1^{Yj}$. The difference between this estimate and the true value of $\beta_1^{Yj}$ is the collider bias induced in the regression coefficient of model (Yj). This bias was plotted against the value of the interaction parameter $\delta_3^{Sk}$ in the correctly specified model (Sk), $k = 1, 2, 3$, to assess their relationship. We then fitted the misspecified log-additive model:
\begin{equation*}
	(S0) \;\;\; : \;\;\; \mathbb{P} (S = 1 | X, Y) \; = \; \exp \left\{ \delta_0^{S0} + \delta_1^{S0} X + \delta_2^{S0} Y + \delta_3^{S0} X Y \right\}
\end{equation*}
and computed the maximum likelihood estimate $\hat{\delta}_3^{S0}$ of the interaction parameter $\delta_3^{S0}$, and then plotted this estimate against the collider bias. The estimate $\hat{\delta}_3^{S0}$ was used in our asymptotic study to approximate the limiting value $\tilde{\delta}_3^{S0}$ of the maximum likelihood estimator; note that $\tilde{\delta}_3^{S0}$ is the value that minimizes the Kullback-Leibler divergence between model (S0) and the true data-generating model for $S$ (which here is model (S1), (S2) or (S3)).

Data were generated and models were fitted using \texttt{R}. The log-additive models were fitted as Poisson regression models, using the \texttt{glm} function.

\subsection{Results}

\begin{figure}[!t]
\centering
\includegraphics[scale = 0.65]{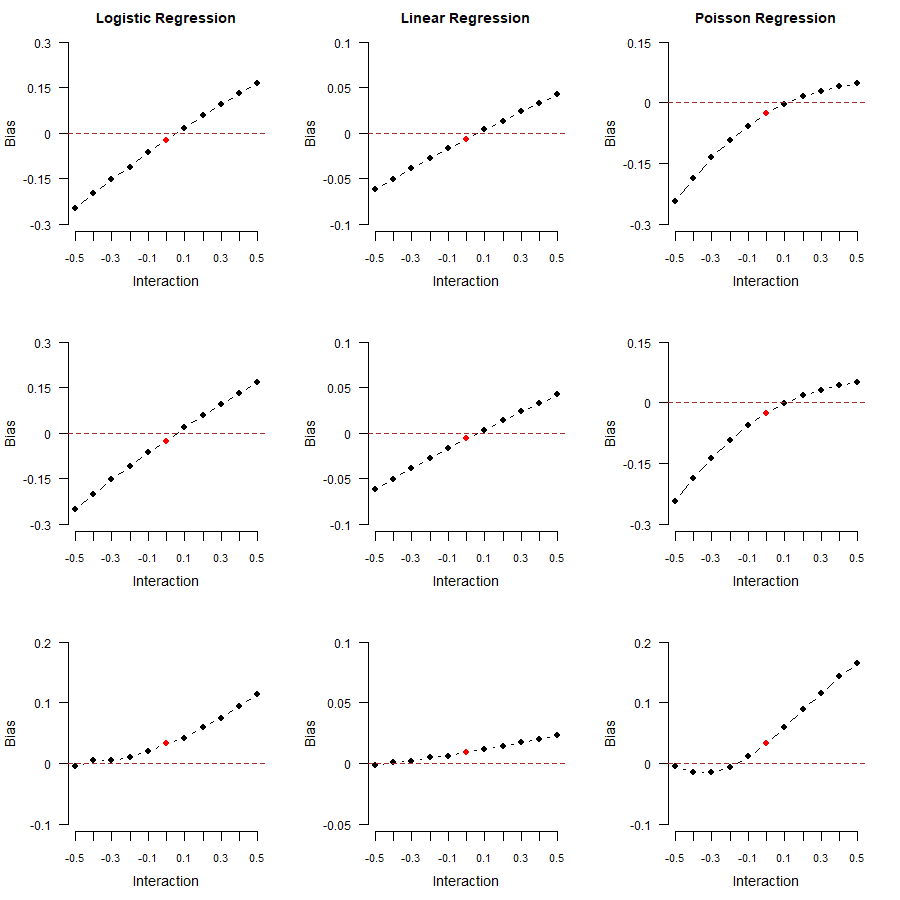}
\caption{Magnitude of collider bias induced in the exposure-outcome regression coefficient by restricting the analysis to selected ($S = 1$) individuals. Data were generated for $n = 10^7$ individuals, and the average selection probability was $50\%$. Outcome data were generated from logistic regression (left column), linear regression (middle column) or Poisson regression (right column) and collider values were generated from logistic regression (model S1, top row), probit regression (model S2, middle row) or the ``double threshold" model (model S3, bottom row). The bias is plotted against the exposure-outcome interaction $\delta_3^{Sk}$ in the collider model. Red color represents simulations with no interaction.} \label{MainSimPlot1m}
\end{figure}

The results of our first asymptotic experiment are shown in Figures~\ref{MainSimPlot1m} and \ref{MainSimPlot1l} and reported in numerical form in Supplementary Section 2. Figure~\ref{MainSimPlot1m} shows the results of simulations where the collider $S$ was generated under a logistic (top row of plots), probit (middle row) or ``double-threshold" (bottom row) model, while the outcome was generated from a logistic (left column), linear (centre column) or Poisson (right column) regression model. The collider bias is plotted against the true value of the interaction parameter $\delta_3^{Sk}$ in the corresponding collider model (Sk), $k = 1, 2, 3$. The relationship between collider bias and the values of the interaction parameters is not linear, with deviations from linearity being more pronounced for Poisson regression and less so for linear regression. In addition, the collider bias is quite small in scenarios where the exposure and outcome do not interact in their effects on the collider ($\delta_3^{Sk} = 0$, plotted in red), although even here some bias still exists.

For some of the models considered here, it is possible to derive analytic expressions for collider bias using arguments similar to those in the previous section. As an example, in Supplementary Section 1.5 we obtain an expression for the collider bias in the exposure-outcome regression coefficient of a logistic regression model when the collider $S$ also follows a logistic regression model. However, such relationships can only be derived for relatively simple models, and the bias will generally depend on all the parameters of the collider model, not just on the interaction term.

\begin{figure}[!t]
\centering
\includegraphics[scale = 0.65]{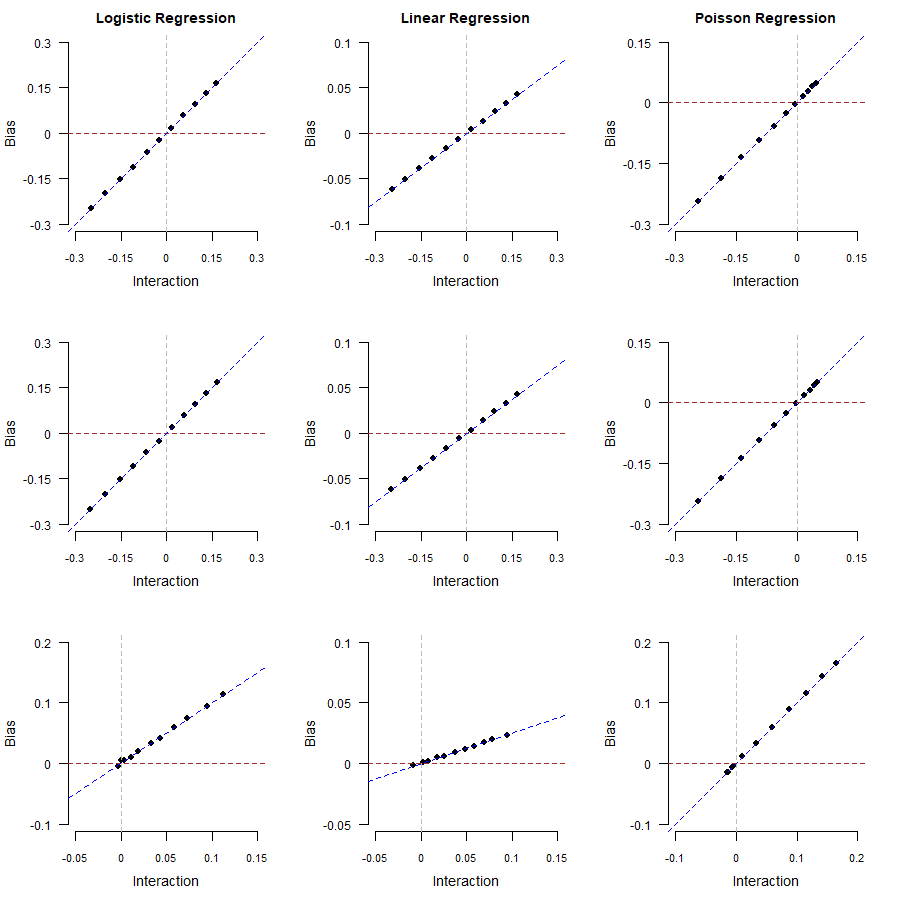}
\caption{Magnitude of collider bias induced in the exposure-outcome regression coefficient by restricting the analysis to selected ($S = 1$) individuals. Data were generated for $n = 10^7$ individuals, and the average selection probability was $50\%$. Outcome data were generated from logistic regression (left column), linear regression (middle column) or Poisson regression (right column) and collider values were generated from logistic regression (model S1, top row), probit regression (model S2, middle row) or the ``double threshold" model (model S3, bottom row). The bias is plotted against the estimated values of the exposure-outcome interaction parameter $\delta_3^{S0}$ in a (misspecified) log-additive model for $S$. A gray vertical line represents no interaction ($\hat{\delta}_3^{S0} = 0$).} \label{MainSimPlot1l}
\end{figure}

In Figure~\ref{MainSimPlot1l}, we plot collider bias against the limiting values $\tilde{\delta}_3^{S0}$ of the log-additive interaction parameter $\delta_3^{S0}$, estimated by fitting the misspecified log-additive model (S0). Note that the points in these plots are not equally spaced along the x-axis because equally spaced interactions on the scale of models (S1)-(S3) do not correspond to equally spaced interactions on the log-additive scale. The relationship between collider bias and limiting values of the estimator of $\delta_3^{S0}$ appears to be linear, with a slope of $1$ for logistic and Poisson regression and $\sigma^2 = 0.25$ in the case of linear regression. This is the same relationship suggested by our theory (equations \eqref{bias.logistic}, \eqref{bias.linear} and \eqref{bias.poisson}), despite the fact that the log-additive model is misspecified.

Note that the linear pattern of collider bias presented in Figure~\ref{MainSimPlot1l} only occurs when the exposure $X$ is binary. In Supplementary Section 3, we report results of an asymptotic study conducted with a normally distributed exposure variable, where the relationship between bias and $\tilde{\delta}_3^{S0}$ is not linear \citep[see also][]{Campbell2005}. Unlike the theoretical results in Section 2 of our manuscript, the distribution of the exposure can affect the magnitude of collider bias when the collider $S$ is not generated from the log-additive model (S0).

\begin{figure}[!t]
\centering
\includegraphics[scale = 0.65]{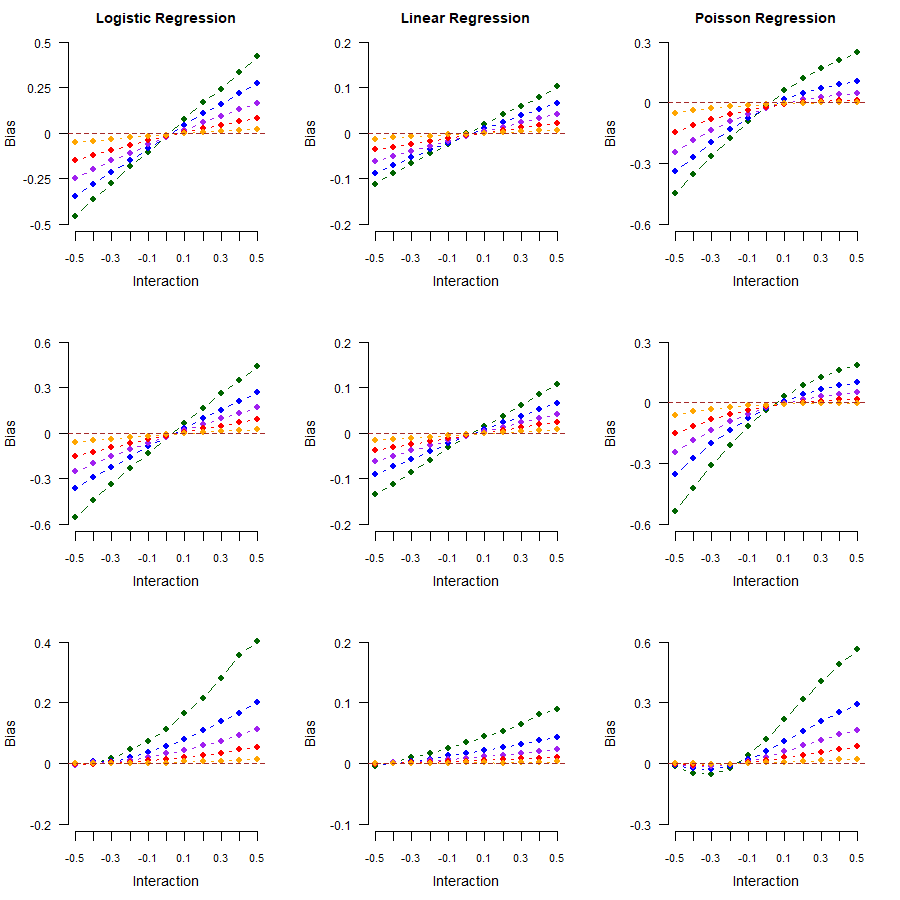}
\caption{Magnitude of collider bias induced in the exposure-outcome regression coefficient by restricting the analysis to selected ($S = 1$) individuals. Different colors represent different selection probabilities (green: $10\%$, blue: $30\%$, purple: $50\%$, red: $70\%$, orange: $90\%$). Outcome data were generated from logistic regression (left column), linear regression (middle column) or Poisson regression (right column) and collider values were generated from logistic regression (model S1, top row), probit regression (model S2, middle row) or the ``double threshold" model (model S3, bottom row). The bias is plotted against the exposure-outcome interaction $\delta_3^{Sk}$ in the collider model.} \label{SupplJointPlot1m}
\end{figure}

\begin{figure}[!t]
\centering
\includegraphics[scale = 0.65]{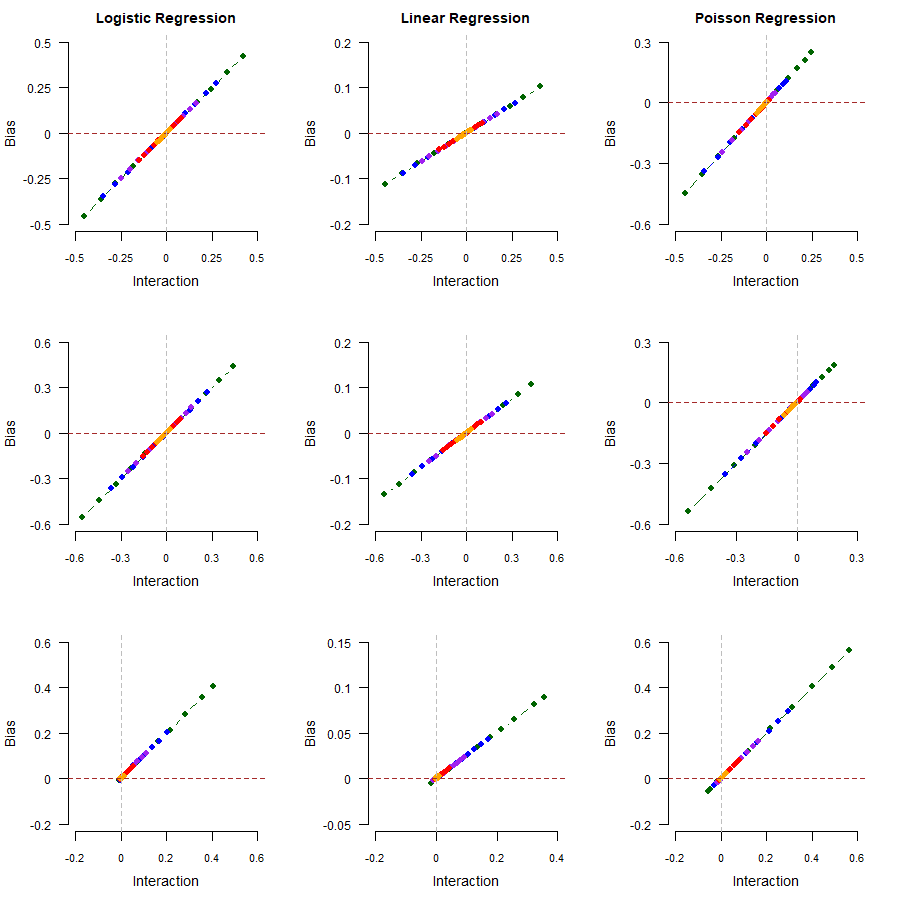}
\caption{Magnitude of collider bias induced in the exposure-outcome regression coefficient by restricting the analysis to selected ($S = 1$) individuals. Different colors represent different selection probabilities (green: $10\%$, blue: $30\%$, purple: $50\%$, red: $70\%$, orange: $90\%$). Outcome data were generated from logistic regression (left column), linear regression (middle column) or Poisson regression (right column) and collider values were generated from logistic regression (model S1, top row), probit regression (model S2, middle row) or the ``double threshold" model (model S3, bottom row). The bias is plotted against estimated values of the exposure-outcome interaction parameter $\delta_3^{S0}$ in a (misspecified) log-additive model for $S$.} \label{SupplJointPlot1l}
\end{figure}

Results from our second asymptotic experiment are shown in Figures~\ref{SupplJointPlot1m} and \ref{SupplJointPlot1l}. In Figure~\ref{SupplJointPlot1m}, we plot the magnitude of collider bias induced in exposure-outcome regression coefficients against the true value of the exposure-outcome interaction in models (S1)-(S3), for a range of selection probabilities: $10\%$ (green), $30\%$ (blue), $50\%$ (purple), $70\%$ (red) and $90\%$ (orange). In Figure~\ref{SupplJointPlot1l}, we do the same for the limiting values $\tilde{\delta}_3^{S0}$ of the interaction parameter $\delta_3^P{S0}$ in the misspecified log-additive model (S0). As seen in Figure~\ref{SupplJointPlot1m}, smaller selection probabilities resulted in more bias across all models considered. However, a smaller proportion of selected individuals also led to a proportional increase in $\tilde{\delta}_3^{S0}$ values. Hence, in Figure~\ref{SupplJointPlot1l}, the relationship between bias and interactions was again linear (with a slope of $1$ for logistic and Poisson-distributed outcomes and $\sigma^2$ for normally distributed outcomes), and the magnitude of collider bias did not depend on the proportion of selected individuals.

Taken together, our results suggest that for binary exposures, the linear relationship between the magnitude of collider bias and the strength of exposure-outcome interaction on the log-additive scale may hold true regardless of whether the log-additive model is a realistic assumption for $S$.

\section{Real-Data Application}

\subsection{Data and Methods}

In addition to the asymptotic study of the previous section, we performed an analysis using data from the Avon Longitudinal Study of Parents and Children \citep[ALSPAC,][]{Boyd2013, Fraser2013}. ALSPAC is a longitudinal population-based study that recruited pregnant women residing in Avon, UK, with expected delivery dates between 1st April 1991 and 31st December 1992. The study included $15447$ pregnancies resulting in $15658$ fetuses, $14901$ of which were alive at 1 year of age. Ethical approval for the study was obtained from the ALSPAC Ethics and Law Committee and the Local Research Ethics Committees. Informed consent for the use of data collected via questionnaires and clinics was obtained from participants following the recommendations of the ALSPAC Ethics and Law Committee at the time. The study website (http://www.bristol.ac.uk/alspac/researchers/our-data/) contains details on all data that is available through a fully searchable data dictionary and variable search tool.

The aim of our analysis is to demonstrate, using real data, the relation between collider bias and exposure-outcome interactions in a log-additive model for selection. To do so, we investigated associations of six maternal traits with child sex. These maternal traits included age at delivery, highest educational qualification held, pre-pregnancy body mass index (BMI), depression status, pre-pregnancy smoking and gestational age. Since child sex is determined randomly at conception and unaffected by environmental exposures, one would expect its associations with maternal traits to be null in the absence of bias \citep[perhaps with the exception of gestational age,][]{Divon2002}. We obtained maternal trait and child sex data for all ALSPAC families; these data were treated as the ``complete sample" for the purposes of our analysis. We also obtained participation data for two follow-up stages of ALSPAC: the ``Teen Focus 4" (TF4) clinic visit (age 17+) and the ``It's All About You" (CCU) questionnaire (age 20). These two subsamples were considered as the ``selected samples" for our analysis. Selection into the two subsamples differed by sex: in TF4, participation rates were $29.5\%$ for males and $40.0\%$ for females, while in the CCU sample, they were $22.0\%$ for males and $35.9\%$ for females. We explored whether these differences could bias estimated associations between maternal traits and offspring sex in the two selected samples compared to the complete sample.

For each of the six maternal traits, we fitted a logistic regression model with child sex as the outcome and the maternal trait as exposure. The models were fitted both in the complete ALSPAC sample and in the two subsamples. We fitted a separate model for each maternal trait to mimic the previous parts of our manuscript, where we only considered one exposure variable. In real-data applications, it may be preferable to conduct a single joint analysis instead, with all six maternal traits included as explanatory variables. Such an analysis is presented in Supplementary Section 5.

We computed regression coefficient estimates from the models regressing child sex on each maternal trait, fitted either to the full ALSPAC sample or to the TF4/CCU subsamples. The difference between estimates in the TF4/CCU subsamples and in the full ALSPAC sample was taken as a measure of collider bias for each trait. We then fitted log-additive models for TF4/CCU participation, each time using child sex and one of the maternal traits as covariates, and compared the interaction estimates in these models to the magnitude of collider bias. In addition, we fitted logistic regression models for TF4 and CCU participation using child sex and one of the six maternal traits as explanatory variables but no interaction terms. Logistic regression is often used to assess which variables associate with study participation, or to adjust for collider bias using inverse probability weighting. We investigated whether parameter estimates from the logistic models could be used to quantify the magnitude of collider bias for each maternal trait.

Note that our application here is conducted for illustrative purposes, to demonstrate the connection between interactions and collider bias in a real dataset. In reality, with access only to the TF4/CCU samples, we would not be able to fit the log-additive model for selection, while with access to the complete ALSPAC sample, there would be no need to restrict the analysis to the TF4/CCU subsamples.

For our analyses, we excluded pregnancies that resulted in miscarriage or early termination ($3.9\%$). We also excluded pregnancies with missing information on maternal traits. Missingness rates for the six maternal traits and sample sizes for our logistic regression analyses are reported in Supplementary Section 4. Missingness in maternal data can be another source of collider bias, and methods such as multiple imputation could be used to adjust for it. In this illustrative application, we choose to ignore this source of bias and focus on the bias induced by restricting to participants in the TF4 and CCU subsamples.

\subsection{Results}

Estimated associations between each maternal trait and child sex from the respective logistic models are reported in Table~\ref{alspac.all}. We report parameter estimates, standard errors and p-values of association between each trait and child sex, obtained either from all ALSPAC participants or only from TF4/CCU attendants.

\begin{center}
\begin{table*}[ht]%
\centering
\scriptsize
\begin{tabular}{c|ccc|ccc|ccc}
  \multirow{2}{*}{Mat Trait} & \multicolumn{3}{c}{All ALSPAC} & \multicolumn{3}{c}{TF4} & \multicolumn{3}{c}{CCU} \\
  & $\hat{\beta}_1$ & s.e. $(\hat{\beta}_1)$ & P-value & $\hat{\beta}_1$ & s.e. $(\hat{\beta}_1)$ & P-value & $\hat{\beta}_1$ & s.e. $(\hat{\beta}_1)$ & P-value \\ 
  \hline
  Age 			&  0.008 & 0.003 & 0.014 &  0.020 & 0.006 & 0.002 &  0.023 & 0.007 & 0.001 \\ 
  Education 	& -0.018 & 0.014 & 0.200 &  0.067 & 0.024 & 0.006 &  0.116 & 0.027 & $1.6 \times 10^{-5}$ \\ 
  BMI 			&  0.004 & 0.005 & 0.369 &  0.010 & 0.008 & 0.245 &  0.007 & 0.009 & 0.431 \\ 
  Depression 	& -0.051 & 0.053 & 0.341 & -0.039 & 0.097 & 0.686 & -0.085 & 0.116 & 0.461 \\ 
  Smoking 		&  0.054 & 0.032 & 0.092 & -0.021 & 0.058 & 0.720 & -0.149 & 0.067 & 0.025 \\ 
  Gest Age 		& -0.033 & 0.007 & $3.3 \times 10^{-6}$ & -0.059 & 0.016 & $1.4 \times 10^{-4}$ & -0.066 & 0.017 & $1.4 \times 10^{-4}$ \\  
  \hline
\end{tabular}
\caption{Associations of maternal traits with child sex in ALSPAC, obtained by fitting six separate logistic regression models for child sex, each with one maternal trait as the exposure. Estimated associations, standard errors and p-values computed using data on either all ALSPAC participants, or only those who attended the TF4 visit, or only those who returned the CCU questionnaire.\label{alspac.all}}
\end{table*}
\end{center}

Mother’s age at delivery and gestational age were associated with child sex in all three samples. The observational association between gestational age and child sex has previously been noted in the literature \citep{Divon2002} and could be due to reverse causation, while the association with mother's age in ALSPAC was fairly weak and could be due to the missing maternal data. Mother’s education was not associated with child sex in the full ALSPAC sample but was seen to associate with child sex in both TF4 and CCU. Maternal smoking was associated with child sex in the CCU sample but not in the TF4 sample or in the full ALSPAC sample, while BMI and depression before pregnancy exhibited no association with child sex in any of the three regression analyses. These results suggest collider bias may be affecting the association of maternal education and smoking with child sex. This is not unreasonable, as smoking and education are often associated with participation in scientific studies, and at the same time, participation rates in TF4 and CCU differed between males and females, as mentioned earlier.

\begin{table*}[ht]%
\centering
\scriptsize
  \begin{tabular}{c|ccc|ccc|ccc|c}
  \multirow{2}{*}{Variable} & \multicolumn{3}{c}{Child Sex} & \multicolumn{3}{c}{Maternal Trait} & \multicolumn{3}{c}{Interaction} & \multirow{2}{*}{Bias} \\
  & Est & StdErr & P-value & Est & StdErr & P-value & Est & StdErr & P-value & \\
  \hline
  \multicolumn{11}{c}{TF4 Estimates} \\
  \hline
  Age 			& -0.539 & 0.171 & 0.002 &  0.046 & 0.004 & $4.5 \times 10^{-33}$ &  0.007 & 0.006 & 0.217 & 0.012 \\ 
  Education 	& -0.583 & 0.086 & $1.0 \times 10^{-11}$ & 0.189 & 0.016 & $3.8 \times 10^{-33}$ &  0.083 & 0.024 & 0.001 & 0.085 \\ 
  BMI 			& -0.399 & 0.186 & 0.032 & -0.008 & 0.005 & 0.149 &  0.005 & 0.008 & 0.562 & 0.006 \\ 
  Depression 	& -0.299 & 0.291 & 0.304 &  0.227 & 0.067 & 0.001 &  0.000 & 0.099 & 0.998 & 0.012 \\ 
  Smoking 		& -0.204 & 0.086 & 0.018 & -0.364 & 0.042 & $6.5 \times 10^{-18}$ & -0.084 & 0.064 & 0.190 & -0.075 \\ 
  Gest Age 		& -0.052 & 0.519 & 0.921 &  0.029 & 0.009 & 0.001 & -0.007 & 0.013 & 0.610 & -0.026 \\ 
  \hline
  \multicolumn{11}{c}{CCU Estimates} \\
  \hline
  Age 			& -0.767 & 0.189 & $4.8 \times 10^{-5}$ &  0.048 & 0.004 & $3.4 \times 10^{-33}$ &  0.009 & 0.006 & 0.176 & 0.015 \\ 
  Education 	& -0.937 & 0.096 & $1.7 \times 10^{-22}$ &  0.190 & 0.016 & $3.5 \times 10^{-31}$ &  0.133 & 0.027 & $5.6 \times 10^{-7}$ & 0.134 \\ 
  BMI 			& -0.526 & 0.208 & 0.011 & -0.012 & 0.006 & 0.031 &  0.002 & 0.009 & 0.807 & 0.003 \\ 
  Depression 	& -0.329 & 0.341 & 0.335 &  0.371 & 0.076 & $1.1 \times 10^{-6}$ & -0.050 & 0.116 & 0.665 & -0.034 \\ 
  Smoking 		& -0.196 & 0.095 & 0.040 & -0.328 & 0.043 & $3.8 \times 10^{-14}$ & -0.238 & 0.072 & 0.001 & -0.203 \\ 
  Gest Age 		& -0.014 & 0.588 & 0.980 &  0.041 & 0.010 & $3.8 \times 10^{-5}$ & -0.012 & 0.015 & 0.411 & -0.033 \\ 
  \hline
\end{tabular}
\caption{Parameter estimates, standard errors and p-values for a log-additive model of TF4 or CCU participation in terms of child sex, maternal traits and interactions between child sex and maternal traits. The observed bias in $\hat{\beta}_1$ estimates calculated from Table~\ref{alspac.all} is also reported for comparison. \label{alspac.logadd}}
\end{table*}

Table~\ref{alspac.logadd} contains the results of fitting log-additive models for TF4 and CCU participation. Again, six different models were fitted, one for each maternal trait. All models also included child sex and an interaction between the maternal trait and child sex. We report parameter estimates, standard errors and $95\%$ p-values for the regression parameters. For comparison, we also report the bias observed in Table~\ref{alspac.all}, computed as the difference between $\beta_1$ estimates in the TF4/CCU samples and in the complete sample.

All maternal traits were associated with CCU participation, and all maternal traits apart from BMI were associated with TF4 participation in their respective models. However, evidence of an interaction between the maternal traits and child sex was observed only for maternal education (in both samples) and smoking (in the CCU sample). This was in line with our previous analysis, in which maternal education associated with child sex among CCU or TF4 participants, and maternal smoking did so among CCU participants. The regression coefficient for the education-child sex interaction was estimated to be positive in both log-additive models; this would suggest positive bias. Indeed the TF4 and CCU regression coefficients in Table~\ref{alspac.all} were both larger than the regression coefficients in the all-ALSPAC analysis (i.e. positive bias). On the other hand, the smoking-child sex interaction in the CCU sample was negative, suggesting negative bias, which was indeed the case based on Table~\ref{alspac.all}. Finally, the interaction parameter estimates were a good approximation of the magnitude of bias caused by restricting to the TF4 or CCU subsamples: for all six traits and for both subsamples, a $95\%$ confidence interval for the interaction parameter in the log-additive model contained the observed value of the bias.

\begin{table*}[ht]%
\centering
\footnotesize
\begin{tabular}{c|ccc|ccc|c}
  \multirow{2}{*}{Variable} & \multicolumn{3}{c}{Child Sex} & \multicolumn{3}{c}{Maternal Trait} & \multirow{2}{*}{Bias}  \\
  & Est & StdErr & P-value & Est & StdErr & P-value & \\ 
  \hline
  \multicolumn{8}{c}{TF4 Estimates} \\
  \hline
  Age 			& -0.524 & 0.036 & $4.6 \times 10^{-47}$ &  0.079 & 0.004 & $8.1 \times 10^{-99}$ & 0.012 \\ 
  Education 	& -0.515 & 0.038 & $1.9 \times 10^{-41}$ &  0.369 & 0.016 & $1.5 \times 10^{-125}$ & 0.085 \\ 
  BMI 			& -0.472 & 0.039 & $1.6 \times 10^{-34}$ & -0.009 & 0.005 & 0.069 & 0.006 \\ 
  Depression 	& -0.482 & 0.037 & $6.4 \times 10^{-38}$ &  0.345 & 0.059 & $5.7 \times 10^{-9}$ & 0.012 \\ 
  Smoking 		& -0.494 & 0.037 & $5.0 \times 10^{-41}$ & -0.603 & 0.039 & $8.3 \times 10^{-55}$ & -0.075 \\ 
  Gest Age 		& -0.483 & 0.036 & $1.1 \times 10^{-41}$ &  0.038 & 0.008 & $9.8 \times 10^{-7}$ & -0.026 \\   
  \hline
  \multicolumn{8}{c}{CCU Estimates} \\
  \hline
  Age 			& -0.752 & 0.038 & $2.8 \times 10^{-85}$ &  0.078 & 0.004 & $5.0 \times 10^{-87}$ & 0.015 \\ 
  Education 	& -0.753 & 0.040 & $4.0 \times 10^{-79}$ &  0.371 & 0.016 & $2.2 \times 10^{-115}$ & 0.134 \\ 
  BMI 			& -0.705 & 0.040 & $3.0 \times 10^{-68}$ & -0.017 & 0.005 & 0.001 & 0.003 \\ 
  Depression 	& -0.705 & 0.039 & $5.8 \times 10^{-72}$ &  0.489 & 0.066 & $1.1 \times 10^{-13}$ & -0.034 \\ 
  Smoking 		& -0.722 & 0.039 & $9.0 \times 10^{-78}$ & -0.594 & 0.041 & $1.3 \times 10^{-47}$ & -0.203 \\ 
  Gest Age 		& -0.704 & 0.038 & $1.1 \times 10^{-77}$ &  0.049 & 0.009 & $9.4 \times 10^{-9}$ & -0.033 \\ 
  \hline
\end{tabular}
\caption{Parameter estimates, standard errors and p-values for a logistic regression of TF4 and CCU participation on child sex and maternal traits. \label{alspac.logit}}
\end{table*}

The results of fitting the logistic regression models without interactions are given in Table~\ref{alspac.logit}. Again, we report parameter estimates, standard errors and p-values for child sex and each maternal variable, as well as the bias observed in ALSPAC. These results suggest strong associations between all six maternal traits and participation in both samples, with the exception of BMI in the TF4 sample. In addition, participation is also associated with child sex, as expected. By fitting the logistic models, an applied researcher could be led to believe that collider bias will occur when studying the associations of maternal traits with child sex in TF4/CCU participants. However, as our analysis in Table~\ref{alspac.all} indicates, bias is present only for maternal education and smoking, and not for the other four maternal traits considered here. This confirms that using a log-additive model with interactions is more informative about collider bias than the commonly used logistic model without interactions.

\section{Discussion}

We have shown that, in three commonly used regression models, the magnitude of collider bias induced in the exposure-outcome association is a linear function of the strength of interaction between the exposure and outcome in a log-additive model for the collider. We have proved these results analytically when the collider is truly distributed according to the log-additive model \eqref{logadd}, and explored them via a numerical asymptotic study and a real-data application in cases when the collider does not follow the log-additive model.


Our results can be useful in several ways. First, modelling selection into a study is an important task for methods that attempt to adjust for collider bias, such as inverse probability weighting (IPW). IPW is typically implemented using logistic regression without interactions as a weighting model; this choice is often made for convenience, and in some applications there is little reason to believe that the logistic model is correctly specified \citep{Seaman2013}. Including interactions in the logistic model can offer more flexibility and better adjustment for collider bias. In addition, our results here raise the question of whether using a log-additive weighting model can improve the performance of IPW compared to the standard logistic model; this is something we will explore in future work.

Second, the fact that collider bias only depends on a single parameter in the simple models considered here can be useful for sensitivity analyses. In some applications, subject-specific knowledge may allow researchers to assess the strength of exposure-outcome interactions, and hence assess whether collider bias is likely to affect their analyses. In addition, simulation studies are sometimes conducted as a form of sensitivity analysis to explore the impact of collider bias in applications. These simulation studies typically work by varying the associations of the exposure, outcome and other relevant variables with study participation and exploring how much collider bias this induces in analysis results. Our work suggests that it is the interactions (on the log-additive scale) that dictate the magnitude of this bias, and therefore that these interactions should also be varied in addition to (or instead of) the exposure-selection and outcome-selection associations.

Third, simulation studies are also used as a tool for assessing the finite-sample performance of novel methods. Our results may therefore be useful to researchers working on methods to detect or adjust for collider bias. For example, it may be desired to design a simulation where complete-case analysis exhibits a specific degree of bias, and compare that with the performance of a newly developed method. This can be done using a log-additive model with an interaction, and specifying the value of the interaction parameter accordingly.

Finally, in some applications it may be possible to fit the log-additive model with interactions to the real data. For example, consider a study where the outcome $Y$ is a disease progression trait. Of interest is the (unconditional) association of an exposure $X$ with the outcome $Y$. Since the progression outcome can only be observed in individuals who have the disease, the study must condition on disease incidence. If the exposure $X$ also associates with disease incidence, the study may suffer from a particular type of collider bias called index event bias \citep{Mitchell2022}: conditioning on incidence will induce associations between $X$ and other common causes of incidence and progression, and hence induce bias in exposure-progression association estimates. If the study has collected information on additional covariates $C$ whose association with incidence is suspected, it is not uncommon to investigate the potential for collider bias by fitting a logistic model for incidence with covariates $X$ and $C$ and no interactions. Our research emphasizes the importance of including interactions in this model, to better assess the potential for collider bias.

Some extensions of our work are possible. Here, we have focused on three simple statistical models for the exposure-outcome association, namely linear, logistic and Poisson regression. It would be interesting to explore whether similar results hold, for example, in survival analysis models. Moreover, extensions of our results to studies of a causal nature are possible. In this manuscript we have compared parameter estimates from conditional and unconditional models. A similar comparison can be made between conditional and unconditional causal effect estimates obtained from causal models, and the role of interactions in the model for $S$ can be explored. This could include instrumental variable analyses, which are known to suffer from collider bias \citep{Gkatzionis2019, Hughes2019}.

We hope our findings will prove useful to methodologists investigating collider bias, as well as to applied researchers attempting to adjust for the bias in their analyses.

\section*{Data Availability Statement}

The \texttt{R} code used to conduct the asymptotic study of our paper is available at the GitHub repository https://github.com/agkatzionis/Interactions-and-collider-bias. Access to ALSPAC data for our real-data application was obtained under application B4189; the data are available upon request to the ALSPAC study: http://www.bristol.ac.uk/alspac/researchers/access/.

\section*{Acknowledgments}

AG and KT received funding for this project by the UK Medical Research Council and the University of Bristol (MRC-IEU core funding, MC UU 00032/02). SRS was funded by UKRI (Unit programme numbers MC UU 00002/10) and was supported by the National Institute for Health Research (NIHR) Cambridge Biomedical Research Centre (BRC-1215-20014). RAH is supported by a Sir Henry Dale Fellowship that is jointly funded by the Wellcome Trust and the Royal Society (grant 215408/Z/19/Z).  The views expressed in this manuscript are those of the authors and not necessarily those of PHE, the NHS, the NIHR or the Department of Health and Social Care.  For the purpose of open access, the authors have applied a Creative Commons Attribution (CC BY) licence to any Author Accepted Manuscript version arising.

Access to ALSPAC data was obtained as part of application B4189. The UK Medical Research Council and Wellcome (Grant ref: 217065/Z/19/Z) and the University of Bristol provide core support for ALSPAC. A comprehensive list of grants funding the ALSPAC study is available on the ALSPAC website (http://www.bristol.ac.uk/alspac/external/documents/grant-acknowledgements.pdf). GWAS data was generated by Sample Logistics and Genotyping Facilities at Wellcome Sanger Institute and LabCorp (Laboratory Corporation of America) using support from 23andMe. We are extremely grateful to all the families who took part in this study, the midwives for their help in recruiting them, and the whole ALSPAC team, which includes interviewers, computer and laboratory technicians, clerical workers, research scientists, volunteers, managers, receptionists and nurses.


\begin{thebibliography}{}

\bibitem[\protect\citeauthoryear{Bartlett, Harel, and Carpenter}{Bartlett
  et~al.}{2015}]{Bartlett2015}
Bartlett, J.~W., O.~Harel, and J.~R. Carpenter (2015, 09).
\newblock Asymptotically unbiased estimation of exposure odds ratios in
  complete records logistic regression.
\newblock {\em American Journal of Epidemiology\/}~{\em 182\/}(8), 730--736.

\bibitem[\protect\citeauthoryear{Berkson}{Berkson}{1946}]{Berkson1946}
Berkson, J. (1946).
\newblock Limitations of the application of fourfold table analysis to hospital
  data.
\newblock {\em Biometrics\/}~{\em 2\/}(3), 47--53.

\bibitem[\protect\citeauthoryear{Boyd, Golding, Macleod, Lawlor, Fraser,
  Henderson, Molloy, Ness, Ring, and Davey~Smith}{Boyd et~al.}{2013}]{Boyd2013}
Boyd, A., J.~Golding, J.~Macleod, D.~A. Lawlor, A.~Fraser, J.~Henderson,
  L.~Molloy, A.~Ness, S.~Ring, and G.~Davey~Smith (2013, 04).
\newblock {Cohort Profile: The ‘Children of the 90s’—the index offspring
  of the Avon Longitudinal Study of Parents and Children}.
\newblock {\em International Journal of Epidemiology\/}~{\em 42\/}(1),
  111--127.

\bibitem[\protect\citeauthoryear{Campbell, Gatto, and Schwartz}{Campbell
  et~al.}{2005}]{Campbell2005}
Campbell, U.~B., N.~M. Gatto, and S.~Schwartz (2005).
\newblock Distributional interaction: Interpretational problems when using
  incidence odds ratios to assess interaction.
\newblock {\em Epidemiologic Perspectives and Innovations\/}~{\em 2\/}(1).

\bibitem[\protect\citeauthoryear{Divon, Ferber, Nisell, and Westgren}{Divon
  et~al.}{2002}]{Divon2002}
Divon, M.~Y., A.~Ferber, H.~Nisell, and M.~Westgren (2002).
\newblock {Male gender predisposes to prolongation of pregnancy}.
\newblock {\em General Obstetrics and Gynecology:
  Fetus-Placenta-Newborn\/}~{\em 187\/}(4), 1081--1083.

\bibitem[\protect\citeauthoryear{Fraser, Macdonald-Wallis, Tilling, Boyd,
  Golding, Davey~Smith, Henderson, Macleod, Molloy, Ness, Ring, Nelson, and
  Lawlor}{Fraser et~al.}{2013}]{Fraser2013}
Fraser, A., C.~Macdonald-Wallis, K.~Tilling, A.~Boyd, J.~Golding,
  G.~Davey~Smith, J.~Henderson, J.~Macleod, L.~Molloy, A.~Ness, S.~Ring, S.~M.
  Nelson, and D.~A. Lawlor (2013, 04).
\newblock {Cohort Profile: The Avon Longitudinal Study of Parents and Children:
  ALSPAC mothers cohort}.
\newblock {\em International Journal of Epidemiology\/}~{\em 42\/}(1), 97--110.

\bibitem[\protect\citeauthoryear{Gkatzionis, Burgess, Conti, and
  Newcombe}{Gkatzionis et~al.}{2020}]{Gkatzionis2019}
Gkatzionis, A., S.~Burgess, D.~V. Conti, and P.~J. Newcombe (2020).
\newblock {Bayesian variable selection with a pleiotropic loss function in
  Mendelian randomization}.
\newblock {\em bioRxiv\/}.

\bibitem[\protect\citeauthoryear{Greenland}{Greenland}{1977}]{Greenland1977}
Greenland, S. (1977).
\newblock Response and follow-up bias in cohort studies.
\newblock {\em American Journal of Epidemiology\/}~{\em 106\/}(3), 184--187.

\bibitem[\protect\citeauthoryear{Greenland}{Greenland}{1996}]{Greenland1996}
Greenland, S. (1996).
\newblock Basic methods for sensitivity analysis of biases.
\newblock {\em International journal of epidemiology\/}~{\em 25\/}(6),
  1107--1116.

\bibitem[\protect\citeauthoryear{Greenland}{Greenland}{2009}]{Greenland2009}
Greenland, S. (2009, 09).
\newblock {Bayesian perspectives for epidemiologic research: III. Bias analysis
  via missing-data methods}.
\newblock {\em International Journal of Epidemiology\/}~{\em 38\/}(6),
  1662--1673.

\bibitem[\protect\citeauthoryear{Hern\'{a}n, Hern\'{a}ndez-Díaz, and
  Robins}{Hern\'{a}n et~al.}{2004}]{Hernan2004}
Hern\'{a}n, M.~A., S.~Hern\'{a}ndez-Díaz, and J.~M. Robins (2004).
\newblock A structural approach to selection bias.
\newblock {\em Epidemiology\/}~{\em 15\/}(5), 615--625.

\bibitem[\protect\citeauthoryear{Hughes, Davies, Davey~Smith, and
  Tilling}{Hughes et~al.}{2019}]{Hughes2019}
Hughes, R.~A., N.~M. Davies, G.~Davey~Smith, and K.~Tilling (2019).
\newblock Selection bias when estimating average treatment effects using
  one-sample instrumental variable analysis.
\newblock {\em Epidemiology\/}~{\em 30\/}(3), 350--357.

\bibitem[\protect\citeauthoryear{Jiang and Ding}{Jiang and
  Ding}{2017}]{Jiang2017}
Jiang, Z. and P.~Ding (2017).
\newblock The directions of selection bias.
\newblock {\em Statistics \& Probability Letters\/}~{\em 125}, 104--109.

\bibitem[\protect\citeauthoryear{Kleinbaum, Kupper, and Morgenstern}{Kleinbaum
  et~al.}{1982}]{Kleinbaum1982}
Kleinbaum, D.~G., L.~L. Kupper, and H.~Morgenstern (1982).
\newblock {\em Epidemiologic research: principles and quantitative methods}.
\newblock John Wiley and Sons, New York.

\bibitem[\protect\citeauthoryear{Mansournia, Nazemipour, and
  Etminan}{Mansournia et~al.}{2022}]{Mansournia2022}
Mansournia, M.~A., M.~Nazemipour, and M.~Etminan (2022, 06).
\newblock {Interaction Contrasts and Collider Bias}.
\newblock {\em American Journal of Epidemiology\/}~{\em 191\/}(10), 1813--1819.

\bibitem[\protect\citeauthoryear{Mitchell, Hartley, Walker, Gkatzionis,
  Yarmolinsky, Bell, Chong, Paternoster, Tilling, and Smith}{Mitchell
  et~al.}{2022}]{Mitchell2022}
Mitchell, R.~E., A.~Hartley, V.~M. Walker, A.~Gkatzionis, J.~Yarmolinsky, J.~A.
  Bell, A.~H.~W. Chong, L.~Paternoster, K.~Tilling, and G.~D. Smith (2022).
\newblock {Strategies to investigate and mitigate collider bias in genetic and
  Mendelian randomization studies of disease progression}.
\newblock {\em medRxiv\/}.

\bibitem[\protect\citeauthoryear{Rothman, Lash, and Greenland}{Rothman
  et~al.}{2008}]{Rothman2008}
Rothman, K.~J., T.~L. Lash, and S.~Greenland (2008).
\newblock {\em Modern Epidemiology\/} (third ed.).
\newblock Lippincott Williams and Wilkins.

\bibitem[\protect\citeauthoryear{Seaman and White}{Seaman and
  White}{2013}]{Seaman2013}
Seaman, S.~R. and I.~R. White (2013).
\newblock Review of inverse probability weighting for dealing with missing
  data.
\newblock {\em Statistical Methods in Medical Research\/}~{\em 22\/}(3),
  278--295.

\bibitem[\protect\citeauthoryear{Shahar and Shahar}{Shahar and
  Shahar}{2017}]{Shahar2017}
Shahar, D.~J. and E.~Shahar (2017).
\newblock A theorem at the core of colliding bias.
\newblock {\em The International Journal of Biostatistics\/}~{\em 13\/}(1),
  20160055.

\bibitem[\protect\citeauthoryear{Sperrin, Candlish, Badrick, Renehan, and
  Buchan}{Sperrin et~al.}{2016}]{Sperrin2016}
Sperrin, M., J.~Candlish, E.~Badrick, A.~Renehan, and I.~Buchan (2016).
\newblock Collider bias is only a partial explanation for the obesity paradox.
\newblock {\em Epidemiology\/}~{\em 27\/}(4), 525--530.

\bibitem[\protect\citeauthoryear{VanderWeele}{VanderWeele}{2015}]{VanderWeele2015}
VanderWeele, T. (2015).
\newblock {\em Explanation in Causal Inference: Methods for Mediation and
  Interaction}.
\newblock Oxford University Press, New York.

\bibitem[\protect\citeauthoryear{Viallon and Dufournet}{Viallon and
  Dufournet}{2016}]{Viallon2016}
Viallon, V. and M.~Dufournet (2016).
\newblock Can collider bias fully explain the obesity paradox?
\newblock arXiv:1612.06547v1.

\bibitem[\protect\citeauthoryear{Wooldridge}{Wooldridge}{2012}]{Wooldridge2012}
Wooldridge, J.~M. (2012).
\newblock {\em Introductory Econometrics: A Modern Approach\/} (Fifth ed.).
\newblock South-western Cengage Learning.

\end{thebibliography}

\end{document}


\title{\textbf{\Large{Relationship between Collider Bias and Interactions on the Log-Additive Scale}}}
\author{Apostolos Gkatzionis$^{1}$, Shaun R. Seaman$^{3}$, Rachael A. Hughes$^{1,2}$, Kate Tilling$^{1,2}$}
\date{\begin{small}$^1$MRC Integrative Epidemiology Unit, University of Bristol, UK. \\
$^2$Population Health Science Institute, Bristol Medical School, University of Bristol, UK. \\
$^3$MRC Biostatistics Unit, University of Cambridge, Cambridge, UK.
\end{small}}
\maketitle

\vspace{1cm}

\begin{center}
\Large{\textbf{Supplementary Material}}
\end{center}

\vspace{1cm}

\section{Derivation of Main Results}

In this section, we derive mathematically the expressions for collider bias presented in the manuscript. The section follows the same structure as in the main part of the manuscript: we consider in turn collider bias in odds ratios for binary outcome variables, collider bias in risk ratios for binary outcome variables; collider bias in linear regression coefficients for continuous outcome variables; and collider bias in Poisson regression coefficients for count outcome variables. Throughout this section, we assume that the collider $S$ is distributed according to the log-additive model
\begin{equation}
	\log \mathbb{P} (S = 1 | X, Y) = \delta_0 + \delta_1 X + \delta_2 Y + \delta_3 X Y \label{logadd.s}
\end{equation}
with the parameter $\delta_3$ expressing the exposure-outcome interaction.

\subsection{Binary Outcome - Collider Bias on the Odds Ratio Scale}

We start with the case of a binary outcome and derive the relationship between the unconditional odds ratio
\begin{equation*}
	OR_{XY} (x) = \frac{\mathbb{P} (Y = 1 | X = x + 1)}{\mathbb{P} (Y = 0 | X = x + 1)} \times \frac{\mathbb{P} (Y = 0 | X = x)}{\mathbb{P} (Y = 1 | X = x)}
\end{equation*}
and the conditional odds ratio
\begin{equation*}
	OR_{XY | S = 1} (x) = \frac{\mathbb{P} (Y = 1 | X = x + 1, S = 1)}{\mathbb{P} (Y = 0 | X = x + 1, S = 1)} \times \frac{\mathbb{P} (Y = 0 | X = x, S = 1)}{\mathbb{P} (Y = 1 | X = x, S = 1)}
\end{equation*}
We note that 
\begin{eqnarray}
	\frac{\mathbb{P} (Y = 1 | X, S = 1)}{\mathbb{P} (Y = 0 | X, S = 1)} & = & \frac{\mathbb{P} (Y = 1 | X)}{\mathbb{P} (Y = 0 | X)} \times \frac{\mathbb{P} (S = 1 | X, Y = 1)}{\mathbb{P} (S = 1 | X, Y = 0)} \nonumber \\
	& = & \frac{\mathbb{P} (Y = 1 | X)}{\mathbb{P} (Y = 0 | X)} \times \frac{\exp \left\{ \delta_0 + \delta_1 X + \delta_2 + \delta_3 X \right\}}{\exp \left\{ \delta_0 + \delta_1 X \right\}} \nonumber \\
	& = & \frac{\mathbb{P} (Y = 1 | X)}{\mathbb{P} (Y = 0 | X)} \exp \left\{ \delta_2 + \delta_3 X \right\} \nonumber
\end{eqnarray}
Therefore, the relationship between the conditional and the unconditional odds ratio is given by
\begin{eqnarray}
	OR_{XY | S = 1} (x) & = & \frac{\mathbb{P} (Y = 1 | X = x + 1, S = 1)}{\mathbb{P} (Y = 0 | X = x + 1, S = 1)} \times \frac{\mathbb{P} (Y = 0 | X = x, S = 1)}{\mathbb{P} (Y = 1 | X = x, S = 1)} \nonumber \\
 	& = & \frac{\mathbb{P} (Y = 1 | X = x + 1)}{\mathbb{P} (Y = 0 | X = x + 1)} \times  \frac{\mathbb{P} (Y = 0 | X = x)}{\mathbb{P} (Y = 1 | X = x)} \times \exp \left\{ \delta_2 + \delta_3 (x + 1) - \delta_2 - \delta_3 x \right\} \nonumber \\
 	& = & OR_{XY} (x) \; \exp \left\{ \delta_3 \right\} \label{bias.or.s}
\end{eqnarray}
which is the result reported in the main part of our manuscript. This result has previously been obtained by several authors in the literature, see e.g. \cite{Jiang2017} and references therein. 

The relationship for logistic regression coefficients follows by noting that if
\begin{equation*}
	\text{logit} \mathbb{P} (Y = 1 | X) = \beta_0 + \beta_1 X
\end{equation*}
and
\begin{equation*}
	\text{logit} \mathbb{P} (Y = 1 | X, S = 1) = \beta_0^S + \beta_1^S X
\end{equation*}
the unconditional odds ratio becomes
\begin{equation*}
	OR_{XY} (x) = \frac{ \frac{e^{\beta_0 + \beta_1 (x + 1)}}{1 + e^{\beta_0 + \beta_1 (x + 1)}} }{ \frac{1}{1 + e^{\beta_0 + \beta_1 (x + 1)}} } \times \frac{ \frac{1}{1 + e^{\beta_0 + \beta_1 x}} }{ \frac{e^{\beta_0 + \beta_1 x}}{1 + e^{\beta_0 + \beta_1 x}} } = e^{\beta_1}
\end{equation*}
and likewise, the conditional odds ratio becomes $OR_{XY | S = 1} (x) = e^{\beta_1^S}$, and then comparing with \eqref{bias.or.s}.

The derivation of similar results for multiple exposures and non-linear exposure-collider effects follows using similar arguments.

\subsection{Binary Outcome - Collider Bias on the Risk Ratio Scale}

We now explore the magnitude of collider bias on the risk ratio scale by comparing the unconditional risk ratio
\begin{equation*}
	RR_{XY} (x) = \frac{\mathbb{P} (Y = 1 | X = x + 1)}{\mathbb{P} (Y = 1 | X = x)}
\end{equation*}
to the conditional risk ratio
\begin{equation*}
	RR_{XY | S = 1} (x) = \frac{\mathbb{P} (Y = 1 | X = x + 1, S = 1)}{\mathbb{P} (Y = 1 | X = x, S = 1)}
\end{equation*}
We have
\begin{eqnarray}
	RR_{XY | S = 1} (x) & = & \frac{ \mathbb{P} (Y = 1, S = 1 | X = x + 1) \;\; / \;\; \mathbb{P} (S = 1 | X = x + 1) }{ \mathbb{P} (Y = 1, S = 1 | X = x) \;\; / \;\; \mathbb{P} (S = 1 | X = x) } \nonumber \\
	& = & \frac{ \mathbb{P} (Y = 1 | X = x + 1) \; \mathbb{P} (S = 1 | X = x + 1, Y = 1) \; \mathbb{P} (S = 1 | X = x) }{ \mathbb{P} (Y = 1 | X = x) \; \mathbb{P} (S = 1 | X = x, Y = 1) \; \mathbb{P} (S = 1 | X = x + 1) } \nonumber \\
	& = & RR_{XY} (x) \times \frac{\mathbb{P} (S = 1 | X = x + 1, Y = 1)}{\mathbb{P} (S = 1 | X = x, Y = 1)} \times \frac{\mathbb{P} (S = 1 | X = x)}{\mathbb{P} (S = 1 | X = x + 1)} \nonumber \\
	& = & RR_{XY} (x) \times \frac{\exp \{ \delta_0 + \delta_1  (x + 1) + \delta_2 + \delta_3 (x + 1) \}}{\exp \{ \delta_0 + \delta_1 x + \delta_2 + \delta_3 x \}} \times \frac{\mathbb{P} (S = 1 | X = x)}{\mathbb{P} (S = 1 | X = x + 1)} \nonumber \\
	& = & RR_{XY} (x) \times \exp \{ \delta_1 + \delta_3 \} \times \frac{\mathbb{P} (S = 1 | X = x)}{\mathbb{P} (S = 1 | X = x + 1)} \nonumber
\end{eqnarray}
In addition,
\begin{eqnarray}
	\mathbb{P} (S = 1 | X = x) & = & \mathbb{P} (S = 1 | X = x, Y = 1) \; \mathbb{P} (Y = 1 | X = x) + \nonumber \\
	& & \hspace{1cm} + \; \mathbb{P} (S = 1 | X = x, Y = 0) \; \mathbb{P} (Y = 0 | X = x) \nonumber \\
	& = & \exp \{ \delta_0 + \delta_1 x + \delta_2 + \delta_3 x \} \; \mathbb{P} (Y = 1 | X = x) + \nonumber \\
	& & \hspace{1cm} + \exp \{ \delta_0 + \delta_1 x \} \; \mathbb{P} (Y = 0 | X = x) \nonumber
\end{eqnarray}
Putting everything together, we have
\begin{eqnarray}
	RR_{XY | S = 1} (x) & = & RR_{XY} (x) \times \exp \{ \delta_1 + \delta_3 \} \times \nonumber \\
	& & \times \frac{ e^{ \delta_0 + \delta_1 x + \delta_2 + \delta_3 x } \; \mathbb{P} (Y = 1 | X = x) + e^{ \delta_0 + \delta_1 x } \; \mathbb{P} (Y = 0 | X = x) }{ e^{ \delta_0 + \delta_1 (x + 1) + \delta_2 + \delta_3 (x + 1) } \; \mathbb{P} (Y = 1 | X = x + 1) + e^{\delta_0 + \delta_1 (x + 1)} \; \mathbb{P} (Y = 0 | X = x + 1) } \nonumber \\
	& = & RR_{XY} (x) \times \frac{ e^{\delta_2 + \delta_3 (x + 1)} \; \mathbb{P} (Y = 1 | X = x) + e^{\delta_3} \; \mathbb{P} (Y = 0 | X = x) }{ e^{ \delta_2 + \delta_3 (x + 1) } \; \mathbb{P} (Y = 1 | X = x + 1) + \mathbb{P} (Y = 0 | X = x + 1) } \label{bias.rr.s}	
\end{eqnarray}

More specific formulae for the bias on  the risk ratio scale can be obtained by making additional modelling assumptions for the exposure-outcome relationship into \eqref{bias.rr.s}. Consider first a log-binomial regression model for $Y$,
\begin{equation*}
	\text{log} \mathbb{P} (Y = 1 | X = x) = \beta_0 + \beta_1 x
\end{equation*}
The regression coefficient $\beta_1$ in this model is known to represent a log-risk ratio:
\begin{equation*}
	RR_{XY} (x) = \frac{\exp \{ \beta_0 + \beta_1 x \}}{\exp \{ \beta_0 \}} = e^{\beta_1}
\end{equation*} 
Moreover, \eqref{bias.rr.s} yields a conditional risk ratio of
\begin{eqnarray}
	RR_{XY | S = 1} (x) & = & e^{\beta_1} \; \frac{ e^{\delta_2 + \delta_3 (x + 1)} \; e^{\beta_0 + \beta_1 x} + e^{\delta_3} \; (1 - e^{\beta_0 + \beta_1 x}) }{ e^{\delta_2 + \delta_3 (x + 1)} \; e^{\beta_0 + \beta_1 (x + 1)} + (1 - e^{\beta_0 + \beta_1 (x + 1)}) } \nonumber \\
	& = & \frac{ e^{\delta_2 + \delta_3 (x + 1) + \beta_0 + \beta_1 (x + 1)} + e^{\delta_3 + \beta_1} \; (1 - e^{\beta_0 + \beta_1 x}) }{ e^{\delta_2 + \delta_3 (x + 1) + \beta_0 + \beta_1 (x + 1)} + (1 - e^{\beta_0 + \beta_1 (x + 1)}) } \nonumber
\end{eqnarray}
This means that the (absolute) bias in estimating risk ratios is
\begin{equation}
	RR_{XY} (x) - RR_{XY | S = 1} (x) = e^{\beta_1} \; \left( 1 - \frac{ e^{\delta_2 + \delta_3 (x + 1) + \beta_0 + \beta_1 x} + e^{\delta_3} \; (1 - e^{\beta_0 + \beta_1 x}) }{ e^{\delta_2 + \delta_3 (x + 1) + \beta_0 + \beta_1 (x + 1)} + (1 - e^{\beta_0 + \beta_1 (x + 1)}) } \right) \label{bias.logit.s}
\end{equation} 
The bias hence depends on both the interaction term $\delta_3$ and the outcome-collider effect parameter $\delta_2$. If $\delta_3 = 0$, \eqref{bias.logit.s} becomes
\begin{equation*}
	RR_{XY} (x) - RR_{XY | S = 1} (x) = e^{\beta_1} \; \left( 1 - \frac{ e^{\delta_2 + \beta_0 + \beta_1 x} + (1 - e^{\beta_0 + \beta_1 x}) }{ e^{\delta_2 + \beta_0 + \beta_1 (x + 1)} + (1 - e^{\beta_0 + \beta_1 (x + 1)}) } \right)
\end{equation*} 
which means that collider bias on the risk ratio scale can arise even in the absence of exposure-outcome interactions. On the other hand, when $\delta_2 = \delta_3 = 0$, we have $RR_{XY} (x) - RR_{XY | S = 1} (x) = 0$, meaning that there is no bias on the risk ratio scale. This is expected since the outcome does not associate with $S$ in that case, hence $S$ is not a collider. 

Moreover, for a null exposure-outcome association ($\beta_1 = 0$), the risk ratio difference \eqref{bias.logit.s} simplifies to
\begin{equation*}
	RR_{XY} (x) - RR_{XY | S = 1} (x) = 1 - \frac{ e^{\delta_2 + \delta_3 (x + 1) + \beta_0} + e^{\delta_3} \; (1 - e^{\beta_0}) }{ e^{\delta_2 + \delta_3 (x + 1) + \beta_0} + (1 - e^{\beta_0}) }
\end{equation*} 
and under both $\beta_1 = 0$ and $\delta_3 = 0$, we obtain again $RR_{XY} (x) - RR_{XY | S = 1} (x) = 0$, which means that in the absence of interactions in model \eqref{logadd.s}, collider bias should not affect a null association on the risk ratio scale.

Alternatively, we can investigate the bias on the risk ratio scale under a logistic regression exposure-outcome model,
\begin{equation*}
	\text{logit} \mathbb{P} (Y = 1 | X = x) = \beta_0 + \beta_1 x
\end{equation*}
This is perhaps less relevant in practice because logistic regression coefficients can be interpreted as log-odds ratios, therefore researchers working with logistic regression models tend to use odds ratios to quantify the exposure-outcome association. Nevertheless, if it is desired to perform inference on the risk ratio scale, one can obtain from \eqref{bias.rr.s} an unconditional risk ratio of
\begin{equation*}
	RR_{XY} (x) = \frac{\text{expit} \{ \beta_0 + \beta_1 x \}}{\text{expit} \{ \beta_0 \}} = \frac{e^{\beta_1} + e^{\beta_0 + \beta_1 (x + 1)}}{1 + e^{\beta_0 + \beta_1 (x + 1)}}
\end{equation*} 
and a conditional risk ratio of
\begin{eqnarray}
	RR_{XY | S = 1} (x) & = & RR_{XY} (x) \; \frac{ e^{\delta_2 + \delta_3 (x + 1)} \; \text{expit} \{ \beta_0 + \beta_1 x \} + e^{\delta_3} \; (1 - \text{expit} \{ \beta_0 + \beta_1 x \}) }{ e^{\delta_2 + \delta_3 (x + 1)} \; \text{expit} \{ \beta_0 + \beta_1 (x + 1) \} + (1 - \text{expit} \{ \beta_0 + \beta_1 (x + 1) \}) } \nonumber \\
	& = & RR_{XY} (x) \; \frac{1 + e^{\beta_0 + \beta_1 (x + 1)}}{1 + e^{\beta_0 + \beta_1 x}} \; \frac{e^{\delta_2 + \delta_3 (x + 1) + \beta_0 + \beta_1 x} + e^{\delta_3}}{e^{\delta_2 + \delta_3 (x + 1) + \beta_0 + \beta_1 (x + 1)} + 1} \nonumber
\end{eqnarray}
Taking their difference yields a bias of
\begin{equation}
	RR_{XY} (x) - RR_{XY | S = 1} (x) = 1 - \frac{1 + e^{\beta_0 + \beta_1 (x + 1)}}{1 + e^{\beta_0 + \beta_1 x}} \; \frac{e^{\delta_2 + \delta_3 (x + 1) + \beta_0 + \beta_1 x} + e^{\delta_3}}{e^{\delta_2 + \delta_3 (x + 1) + \beta_0 + \beta_1 (x + 1)} + 1} \label{bias.rr.logit.s}
\end{equation} 
Again, the bias depends on both $\delta_2$ and $\delta_3$. If $\delta_3 = 0$, the bias is still present and is equal to
\begin{equation*}
	RR_{XY} (x) - RR_{XY | S = 1} (x) = 1 - \frac{1 + e^{\beta_0 + \beta_1 (x + 1)}}{1 + e^{\beta_0 + \beta_1 x}} \; \frac{1 + e^{\delta_2 + \beta_0 + \beta_1 x}}{1 + e^{\delta_2 + \beta_0 + \beta_1 (x + 1)}}
\end{equation*} 
However, when $\delta_2 = \delta_3 = 0$, the conditional and unconditional risk ratios are equal and the bias is eliminated. Likewise, when $\beta_1 = 0$, the bias simplifies to
\begin{equation*}
	RR_{XY} (x) - RR_{XY | S = 1} (x) = 1 - \frac{e^{\delta_2 + \delta_3 (x + 1) + \beta_0} + e^{\delta_3}}{e^{\delta_2 + \delta_3 (x + 1) + \beta_0} + 1}
\end{equation*} 
and under both $\beta_1 = 0$ and $\delta_3 = 0$, the bias is again null.

\subsection{Continuous Outcome - Collider bias in Linear Regression Coefficients}

We now turn our attention to continuous outcome variables, and assume that the outcome is distributed according to the linear regression model
\begin{equation*}
	Y | X = \beta_0 + \beta_1 X + \epsilon_Y \;\;\; , \;\;\; \epsilon_Y \sim N(0, \sigma^2)
\end{equation*}
where $\mathbb{E} (Y | X) = \beta_0 + \beta_1 X$. Our aim is to explore the bias induced in $\beta_1$ by conditioning on $S = 1$. To achieve this, we will derive an expression for $\mathbb{E} (Y | X, S = 1)$ from first principles. We have
\begin{eqnarray}
	\mathbb{E} (Y | X = x, S = 1) & = & \int y f_Y (y | X = x, S = 1) dy \nonumber \\
	& = & \int y \frac{f_{Y, S} (y, 1 | X = x)}{f_S (1 | X = x)} dy \nonumber \\
	& = & \frac{\int y f_{Y, S} (y, 1 | X = x) dy}{f_S (1 | X = x)} \nonumber \\
	& = & \frac{\int y f_{Y, S} (y, 1 | X = x) dy}{\int f_{Y, S} (\psi, 1 | X = x) d\psi} \nonumber \\
	& = & \frac{\int y f_S (1 | X = x, Y = y) f_Y (y | X = x) dy}{\int f_S (1 | X = x, Y = \psi) f_Y (\psi | X = x) d\psi} \nonumber
\end{eqnarray}
Now use the distributional assumptions for $Y, S$ to get
\begin{eqnarray}
	\mathbb{E} (Y | X = x, S = 1) & = & \frac{\int y \exp \{ \delta_0 + \delta_1 x + \delta_2 y + \delta_3 x y \} \; \frac{1}{\sqrt{2 \pi \sigma^2}} \exp \left\{ - \frac{1}{2 \sigma^2} (y - \beta_0 - \beta_1 x)^2 \right\} dy}{\int \exp \{ \delta_0 + \delta_1 x + \delta_2 \psi + \delta_3 x \psi \} \; \frac{1}{\sqrt{2 \pi \sigma^2}} \exp \left\{ - \frac{1}{2 \sigma^2} (\psi - \beta_0 - \beta_1 x)^2 \right\} d\psi} \nonumber \\
	& = & \frac{\int y \exp \left\{ \delta_2 y + \delta_3 x y - \frac{1}{2 \sigma^2} (y - \beta_0 - \beta_1 x)^2 \right\} dy}{\int \exp \left\{ \delta_2 \psi + \delta_3 x \psi - \frac{1}{2 \sigma^2} (\psi - \beta_0 - \beta_1 x)^2 \right\} d\psi} \label{linear.cond.s}
\end{eqnarray}
The integral in the numerator can be computed by completing the square in the exponent:
\begin{eqnarray}
	N & = & \delta_2 y + \delta_3 x y - \frac{1}{2 \sigma^2} (y - \beta_0 - \beta_1 x)^2 \nonumber \\
	& = & - \frac{1}{2 \sigma^2} \; \left[ (y - \beta_0 - \beta_1 x)^2 - 2 \sigma^2 (\delta_2 y + \delta_3 x y) \right] \nonumber \\
	& = & - \frac{1}{2 \sigma^2} \; \left[ y^2 - 2 y (\beta_0 + \beta_1 x + \sigma^2 \delta_2 + \sigma^2 \delta_3 x) + (\beta_0 + \beta_1 x)^2 \right] \nonumber \\
	& = & - \frac{1}{2 \sigma^2} \; \left[ y^2 - 2 y C_1 + C_2 \right] \nonumber \\
	& = & - \frac{1}{2 \sigma^2} \; \left[ y^2 - 2 y C_1 + C_1^2 - C_1^2 + C_2 \right] \nonumber \\
	& = & - \frac{1}{2 \sigma^2} \; \left[ (y - C_1)^2 \right] + \frac{1}{2 \sigma^2} (C_1^2 - C_2) \nonumber
\end{eqnarray}
where $C_1 = C_1 (x) = \beta_0 + \beta_1 x + \sigma^2 \delta_2 + \sigma^2 \delta_3 x$ and $C_2 = C_2 (x) = (\beta_0 + \beta_1 x)^2$. Equation \eqref{linear.cond.s} then becomes
\begin{eqnarray}
	\mathbb{E} (Y | X = x, S = 1) & = & \frac{\int y \exp \left\{ - \frac{1}{2 \sigma^2} \; (y - C_1)^2 + \frac{1}{2 \sigma^2} (C_1^2 - C_2) \right\} dy}{\int \exp \left\{ - \frac{1}{2 \sigma^2} \; (\psi - C_1)^2 + \frac{1}{2 \sigma^2} (C_1^2 - C_2) \right\} d\psi} \nonumber \\
	& = & \frac{\frac{1}{\sqrt{2 \pi \sigma^2}} \int y \exp \left\{ - \frac{1}{2 \sigma^2} \; (y - C_1)^2 \right\} dy}{\frac{1}{\sqrt{2 \pi \sigma^2}} \int \exp \left\{ - \frac{1}{2 \sigma^2} \; (\psi - C_1)^2 \right\} d\psi} \nonumber
\end{eqnarray}
The denominator is the integral of a $N(C_1, \sigma^2)$ density across its support, therefore equal to 1, while the numerator is the mean value of such a density, which yields
\begin{eqnarray}
	\mathbb{E} (Y | X = x, S = 1) & = & \frac{1}{\sqrt{2 \pi \sigma^2}} \int y \exp \left\{ - \frac{1}{2 \sigma^2} \; (y - C_1)^2 \right\} dy \nonumber \\
	& = & C_1 (x) \;\; = \;\; (\beta_0 \delta_2 \sigma^2) + (\beta_1 + \delta_3 \sigma^2) x \nonumber
\end{eqnarray}
At the same time, if $\beta_0^S$, $\beta_1^S$ denote the regression coefficients of a linear regression model conditional on $S = 1$, we have
\begin{equation*}
	\mathbb{E} (Y | X = x, S = 1) = \beta_0^S + \beta_1^S x
\end{equation*}
and comparing the two expressions for the conditional expectation, we obtain
\begin{eqnarray}
	\beta_0^S & = & \beta_0 + \delta_2 \sigma^2 \nonumber \\
	\beta_1^S & = & \beta_1 + \delta_3 \sigma^2 \nonumber
\end{eqnarray}
The proofs for multiple exposure variables and non-linear exposure-outcome associations follow in a similar way.

\subsection{Count Outcome - Collider bias in Poisson Regression Coefficients}

We now discuss the case of a count outcome variable.  Assume that the relationship between $X$ and $Y$ is characterized by the Poisson regression:
\begin{equation}
	Y | X \sim Poisson (\lambda) \;\;\; , \;\;\; \lambda = \lambda (X) = \exp \left\{ \beta_0 + \beta_1 X \right\} \label{poisson.model.s}
\end{equation}
and that, as earlier, the exposure and the outcome affect $S$ on a log-additive scale. The probability mass function of the Poisson-distributed random variable $Y | X = x$ is:
\begin{eqnarray}
	\mathbb{P} (Y = y | X = x) & = & \frac{\exp \{- \lambda (x)\} (\lambda (x))^{y}}{y!} \nonumber \\
	& = & \frac{1}{y!} \exp \left\{ - e^{\beta_0 + \beta_1 x} \right\} \left( e^{\beta_0 + \beta_1 x} \right)^y \nonumber \\
	& = & \frac{1}{y!} \exp \left\{ y (\beta_0 + \beta_1 x) - e^{\beta_0 + \beta_1 x} \right\} \nonumber
\end{eqnarray}
For the conditional model $Y| X = x, S = 1$, we have
\begin{eqnarray}
	\mathbb{P} (Y = y | X = x, S = 1) & = & \frac{\mathbb{P} (Y = y, S = 1 | X = x)}{\mathbb{P} (S = 1 | X = x} \nonumber \\
	& = & \frac{\mathbb{P} (S = 1 | Y = y, X = x) \; \mathbb{P} (Y = y | X = x)}{\sum_\psi \mathbb{P} (S = 1 | Y = \psi, X = x) \mathbb{P} (Y = \psi | X = x)} \nonumber \\
	& = & \frac{\exp \left\{ \delta_0 + \delta_1 x + \delta_2 y + \delta_3 x y \right\} \frac{1}{y!} \exp \left\{ y (\beta_0 + \beta_1 x) - e^{\beta_0 + \beta_1 x} \right\}}{\sum_\psi \exp \left\{ \delta_0 + \delta_1 x + \delta_2 \psi + \delta_3 x \psi \right\} \frac{1}{\psi!} \exp \left\{ \psi (\beta_0 + \beta_1 x) - e^{\beta_0 + \beta_1 x} \right\}} \nonumber \\
	& = & \frac{\exp \left\{ \delta_2 y + \delta_3 x y \right\} \frac{1}{y!} \exp \left\{ y (\beta_0 + \beta_1 x) \right\}}{\sum_\psi \exp \left\{ \delta_2 \psi + \delta_3 x \psi \right\} \frac{1}{\psi!} \exp \left\{ \psi (\beta_0 + \beta_1 x) \right\}} \nonumber \\
	& = & \frac{\frac{1}{y!} \exp \left\{ y ( (\delta_2 + \delta_3 x) + (\beta_0 + \beta_1 x) ) \right\}}{\sum_\psi \frac{1}{\psi!} \exp \left\{ \psi ( (\delta_2 + \delta_3 x) + (\beta_0 + \beta_1 x) ) \right\}} \nonumber \\
	& = & \frac{\frac{1}{y!} \left\{ \exp \left\{ (\delta_2 + \delta_3 x) + (\beta_0 + \beta_1 x) \right\} \right\}^y}{\sum_\psi \frac{1}{\psi!} \left\{ \exp \left\{ (\delta_2 + \delta_3 x) + (\beta_0 + \beta_1 x) \right\} \right\}^\psi} \nonumber
\end{eqnarray}
Setting $\kappa(x) = \exp \left\{ (\delta_2 + \delta_3 x) + (\beta_0 + \beta_1 x) \right\}$, the numerator and denominator can be recognized as probability mass functions of a Poisson distribution with parameter $\kappa(x)$:
\begin{eqnarray}
	\mathbb{P} (Y = y | X = x, S = 1) & = & \frac{\frac{1}{y!} \left\{ \kappa(x) \right\}^y}{\sum_\psi \frac{1}{\psi!} \left\{ \kappa(x) \right\}^\psi} \nonumber \\
	& = & \frac{\frac{1}{y!} \exp \{- \kappa(x)\} \left\{ \kappa(x) \right\}^y}{\sum_\psi \frac{1}{\psi!} \exp \{- \kappa(x)\} \left\{ \kappa(x) \right\}^\psi} \nonumber \\
	& = & \frac{1}{y!} \exp \{- \kappa(x)\} \left\{ \kappa(x) \right\}^y \nonumber
\end{eqnarray}
where the denominator is equal to $1$ because we are summing over the probability mass function's support. Denoting $\beta_0^S$, $\beta_1^S$ the regression coefficients of a Poisson regression model conditioned on $S = 1$ and comparing the conditional and unconditional exposure-outcome associations, we have
\begin{eqnarray}
	\mathbb{E} (Y | X = x) & = & \exp \left\{ \beta_0 + \beta_1 x \right\} \nonumber \\
	\mathbb{E} (Y | X = x, S = 1) & = & \exp \left\{ \beta_0^S + \beta_1^S x \right\} \nonumber \\
	& = & \exp \left\{ (\beta_0 + \delta_2) + (\beta_1 + \delta_3) x \right\} \nonumber
\end{eqnarray}
from which it readily follows that $\beta_0^S = \beta_0 + \delta_2$ and $\beta_1^S = \beta_1 + \delta_3$.

The result can be extended to Poisson regression models with multiple exposure variables, or to models with a non-linear exposure-collider association. The proofs for these results are similar to the one given above.

\subsection{Collider Bias under a Logistic Model for the Collider}

Similar arguments as in the previous sections can be used to obtain expressions for collider bias when the collider $S$ does not follow the log-additive model \eqref{logadd.s}. Here, we provide an indicative example by computing an expression for the collider bias in the logistic regression coefficient $\beta_1$ of a binary outcome:
\begin{equation*}
	\text{logit} \mathbb{P} (Y = 1 | X) = \beta_0 + \beta_1 X
\end{equation*}
when the collider also follows a logistic model:
\begin{equation}
	\text{logit} \mathbb{P} (S = 1 | X, Y) = \delta_0 + \delta_1 X + \delta_2 Y + \delta_3 X Y \label{logistic.s}
\end{equation}
Similar to our proof for collider bias on the odds ratio scale under a log-additive model, we have
\begin{eqnarray}
	\frac{\mathbb{P} (Y = 1 | X, S = 1)}{\mathbb{P} (Y = 0 | X, S = 1)} & = & \frac{\mathbb{P} (Y = 1 | X)}{\mathbb{P} (Y = 0 | X)} \times \frac{\mathbb{P} (S = 1 | X, Y = 1)}{\mathbb{P} (S = 1 | X, Y = 0)} \nonumber \\
	& = & \frac{\mathbb{P} (Y = 1 | X)}{\mathbb{P} (Y = 0 | X)} \times \frac{\text{expit} \left\{ \delta_0 + \delta_1 X + \delta_2 + \delta_3 X \right\}}{\text{expit} \left\{ \delta_0 + \delta_1 X \right\}} \nonumber \\
	& = & \frac{\mathbb{P} (Y = 1 | X)}{\mathbb{P} (Y = 0 | X)} \times \frac{ \frac{ \exp \{\delta_0 + \delta_1 X + \delta_2 + \delta_3 X \} }{1 + \exp \{\delta_0 + \delta_1 X + \delta_2 + \delta_3 X \} } }{ \frac{ \exp \{\delta_0 + \delta_1 X \} }{1 + \exp \{\delta_0 + \delta_1 X \} } } \nonumber \\
	& = & \frac{\mathbb{P} (Y = 1 | X)}{\mathbb{P} (Y = 0 | X)} \times \frac{ e^{\delta_2 + \delta_3 X} \; \left(1 + e^{\delta_0 + \delta_1 X} \right) }{ 1 + e^{\delta_0 + \delta_1 X + \delta_2 + \delta_3 X} } \nonumber
\end{eqnarray}
Under a logistic regression model for $Y$, the unconditional odds ratio is $OR_{XY} (x) = e^{\beta_1}$ and the conditional odds ratio is
\begin{eqnarray}
	OR_{XY | S = 1} (x) & = & OR_{XY} (x) \; \frac{ \frac{ e^{\delta_2 + \delta_3 (x + 1)} \; \left( 1 + e^{\delta_0 + \delta_1 (x + 1)} \right) }{ 1 + e^{\delta_0 + \delta_1 (x + 1) + \delta_2 + \delta_3 (x + 1)} } }{ \frac{ e^{\delta_2 + \delta_3 x} \; \left( 1 + e^{\delta_0 + \delta_1 x} \right) }{ 1 + e^{\delta_0 + \delta_1 x + \delta_2 + \delta_3 x} } } \nonumber \\
	& = & e^{\beta_1} \; e^{\delta_3} \; \frac{ (1 + e^{\delta_0 + \delta_1 (x + 1)}) \; (1 + e^{\delta_0 + \delta_1 x + \delta_2 + \delta_3 x}) }{ (1 + e^{\delta_0 + \delta_1 x}) \; (1 + e^{\delta_0 + \delta_1 (x + 1) + \delta_2 + \delta_3 (x + 1)}) } \nonumber
\end{eqnarray}
Since regression coefficients in logistic regression models represent log-odds ratios, the magnitude of collider bias in the $X-Y$ regression coefficient is equal to
\begin{eqnarray}
	\log OR_{XY | S = 1} (x) - \log OR_{XY} (x) & = & \delta_3 + \log \frac{ (1 + e^{\delta_0 + \delta_1 (x + 1)}) \; (1 + e^{\delta_0 + \delta_1 x + \delta_2 + \delta_3 x}) }{ (1 + e^{\delta_0 + \delta_1 x}) \; (1 + e^{\delta_0 + \delta_1 (x + 1) + \delta_2 + \delta_3 (x + 1)}) } \nonumber \\
	& = & \delta_3 + \log \left(1 + e^{\delta_0 + \delta_1 (x + 1)} \right) + \log \left(1 + e^{\delta_0 + \delta_1 x + \delta_2 + \delta_3 x} \right) \nonumber \\
	& & \hspace{0.5cm} - \log \left(1 + e^{\delta_0 + \delta_1 x} \right) - \log \left(1 + e^{\delta_0 + \delta_1 (x + 1) + \delta_2 + \delta_3 (x + 1)} \right) \nonumber
\end{eqnarray}
Note that expression depends on all four parameters $\delta_0, \delta_1, \delta_2, \delta_3$ in the collider model.

The appeal of the log-additive model \eqref{logadd.s} over the logistic collider model considered here lies in its simplicity; the fact that collider bias is a linear function of a single parameter, and that this linear relationship holds across a range of outcome models, make the log-additive model a suitable framework for quantifying collider bias in applied studies and designing sensitivity analyses. This would be difficult under the logistic model \eqref{logistic.s} and even harder for more complex models, for which analytic expressions for the bias may not be available.

\section{Numerical Results for the Paper's Asymptotic Study}

In this section we provide numerical results for the asymptotic study conducted in the main part of the paper. We focus on the first asymptotic experiment, where $50\%$ of individuals were selected into the study, the results of which were plotted in Figures~2 and 3 of the main part of the paper. For each of the nine simulation scenarios we report the magnitude of collider bias induced in the exposure-outcome association, along with the true value of the exposure-outcome interaction used to simulate the data in models (S1)-(S3) and the estimated value of the exposure-outcome interaction obtained by fitting the misspecified log-additive model (S0). These are reported in Table~\ref{Suppl.table2}. The results confirm the close relationship between collider bias and estimated exposure-outcome interactions on the log-additive scale: the largest difference between the two among our $9$ scenarios and $11$ interaction values considered in each scenario was $0.003$.

\begin{table*}[!t]
\centering
\footnotesize
\begin{tabular}{c|cc|cc|cc}
  \hline
  \multirow{2}{*}{$\delta_3^{Sk}$} & \multicolumn{2}{c}{Binary Outcome} & \multicolumn{2}{c}{Continuous Outcome} & \multicolumn{2}{c}{Count Outcome} \\
 & Bias & $\hat{\delta}_3^{S0}$ & Bias & $\sigma^2 \; \hat{\delta}_3^{S0}$ & Bias & $\hat{\delta}_3^{S0}$ \\
  \hline
    \multicolumn{7}{c}{Collider Model (S1) - Logistic Regression} \\
  \hline  
  -0.5 & -0.248 & -0.248 & -0.061 & -0.061 & -0.243 & -0.243 \\ 
  -0.4 & -0.197 & -0.200 & -0.050 & -0.050 & -0.187 & -0.187 \\ 
  -0.3 & -0.151 & -0.152 & -0.039 & -0.039 & -0.135 & -0.137 \\ 
  -0.2 & -0.111 & -0.110 & -0.028 & -0.028 & -0.092 & -0.092 \\ 
  -0.1 & -0.063 & -0.063 & -0.017 & -0.017 & -0.057 & -0.056 \\ 
    0  & -0.022 & -0.022 & -0.007 & -0.006 & -0.027 & -0.026 \\ 
   0.1 &  0.017 &  0.016 &  0.004 &  0.004 & -0.003 & -0.003 \\ 
   0.2 &  0.058 &  0.057 &  0.013 &  0.014 &  0.015 &  0.015 \\ 
   0.3 &  0.095 &  0.094 &  0.024 &  0.023 &  0.029 &  0.029 \\ 
   0.4 &  0.130 &  0.130 &  0.033 &  0.033 &  0.040 &  0.039 \\ 
   0.5 &  0.164 &  0.164 &  0.042 &  0.042 &  0.048 &  0.047 \\ 
  \hline  
    \multicolumn{7}{c}{Collider Model (S2) - Probit Regression} \\
  \hline  
  -0.5 & -0.250 & -0.249 & -0.062 & -0.062 & -0.243 & -0.244 \\ 
  -0.4 & -0.201 & -0.201 & -0.050 & -0.050 & -0.185 & -0.185 \\ 
  -0.3 & -0.152 & -0.151 & -0.038 & -0.038 & -0.138 & -0.138 \\ 
  -0.2 & -0.108 & -0.108 & -0.028 & -0.028 & -0.093 & -0.092 \\ 
  -0.1 & -0.065 & -0.066 & -0.016 & -0.017 & -0.056 & -0.056 \\ 
    0  & -0.026 & -0.025 & -0.006 & -0.006 & -0.025 & -0.025 \\ 
   0.1 &  0.020 &  0.018 &  0.003 &  0.004 & -0.001 & -0.001 \\ 
   0.2 &  0.058 &  0.058 &  0.014 &  0.014 &  0.018 &  0.018 \\ 
   0.3 &  0.095 &  0.096 &  0.024 &  0.023 &  0.032 &  0.032 \\ 
   0.4 &  0.130 &  0.131 &  0.032 &  0.033 &  0.044 &  0.043 \\ 
   0.5 &  0.169 &  0.166 &  0.042 &  0.042 &  0.050 &  0.050 \\ 
  \hline  
    \multicolumn{7}{c}{Collider Model (S3) - ``Double Threshold"} \\
  \hline
  -0.5 & -0.004 & -0.003 & -0.002 & -0.002 & -0.005 & -0.004 \\ 
  -0.4 &  0.005 &  0.001 &  0.001 &  0.001 & -0.014 & -0.013 \\ 
  -0.3 &  0.005 &  0.004 &  0.002 &  0.002 & -0.015 & -0.015 \\ 
  -0.2 &  0.011 &  0.011 &  0.005 &  0.005 & -0.006 & -0.006 \\ 
  -0.1 &  0.021 &  0.019 &  0.006 &  0.007 &  0.011 &  0.010 \\ 
    0  &  0.033 &  0.034 &  0.009 &  0.010 &  0.033 &  0.034 \\ 
   0.1 &  0.042 &  0.043 &  0.012 &  0.012 &  0.060 &  0.060 \\ 
   0.2 &  0.060 &  0.059 &  0.014 &  0.015 &  0.089 &  0.088 \\ 
   0.3 &  0.075 &  0.073 &  0.017 &  0.017 &  0.115 &  0.115 \\ 
   0.4 &  0.094 &  0.094 &  0.020 &  0.020 &  0.143 &  0.142 \\ 
   0.5 &  0.113 &  0.113 &  0.023 &  0.024 &  0.166 &  0.165 \\ 
   \hline
\end{tabular}
\caption{Magnitude of collider bias induced in the exposure-outcome regression coefficient by restricting the analysis to selected ($S = 1$) individuals, in the scenario with a binary exposure and an average selection probability of $50\%$. Outcome data are generated from logistic regression (``Binary Outcome" - left), linear regression (``Continuous Outcome" - middle) or Poisson regression (``Count Outcome" - right) and collider values are generated from models (S1)-(S3) as indicated. We report the true value of the exposure-outcome interaction parameter used to generate the data in each of the three collider models (``$\delta_3^{Sk}$"), along with the observed bias in exposure-outcome regression coefficients (``Bias") and estimates of the exposure-outcome interaction parameter in the log-additive model (S0) (``$\hat{\delta}_3^{S1}$"). \label{Suppl.table2}}
\end{table*}

\section{Asymptotic Results with a Continuous Exposure Variable}

\begin{figure}[!t]
\centering
\includegraphics[scale = 0.65]{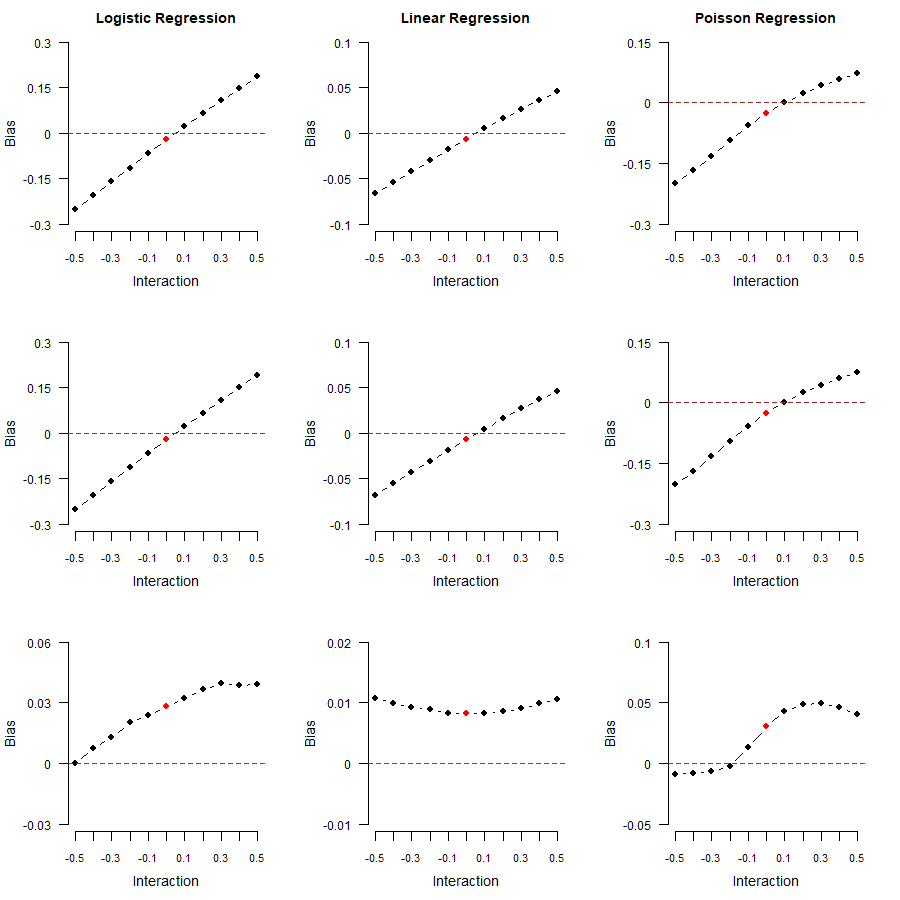}
\caption{Magnitude of collider bias induced in the exposure-outcome regression coefficient by restricting the analysis to selected ($S = 1$) individuals. Exposure data were generated as a continuous variable ($X \sim N(0, 1)$). Outcome data were generated from logistic regression (left column), linear regression (middle column) or Poisson regression (right column) and collider values were generated from logistic regression (model S1, top row), probit regression (model S2, middle row) or the ``double threshold" model (model S3, bottom row). The bias is plotted against the exposure-outcome interaction $\delta_3^{Sk}$ in the collider model. Red color represents simulations with no interaction.} \label{SupplAsyPlot4m}
\end{figure}

So far, we have explored the relationship between collider bias in models (Y1)-(Y3) and exposure-outcome interactions in the log-additive model \eqref{logadd.s} for $S$. We have shown theoretically that when model \eqref{logadd.s} is correctly specified, the relationship is linear, and that holds true regardless of the type of the exposure variable. We have also shown numerically that the relationship appears to be linear when model \eqref{logadd.s} is misspecified and the exposure is a binary variable. However, the same does not hold when the log-additive model is misspecified and the exposure is a continuous variable. In this section, we extend our asymptotic study to demonstrate this fact.

We repeated our first asymptotic experiment, where $50\%$ of simulated individuals are selected ($S = 1$), but generated exposure values from a $N(0, 1)$ distribution. All nine combinations of outcome models and selection models were considered. Once again, we first plotted collider bias against $X-Y$ interactions on the scale of the collider model used to generate the data (logistic, probit or ``double-threshold"); these are visualized in Figure~\ref{SupplAsyPlot4m}. The pattern of bias is somewhat different to that in Figure 2 of the main paper for the ``double-threshold" model, but otherwise the two figures lead to similar conclusions: the strength of the exposure-outcome interaction affects the magnitude of bias, but their relationship is not easy to characterize.

\begin{figure}[!t]
\centering
\includegraphics[scale = 0.65]{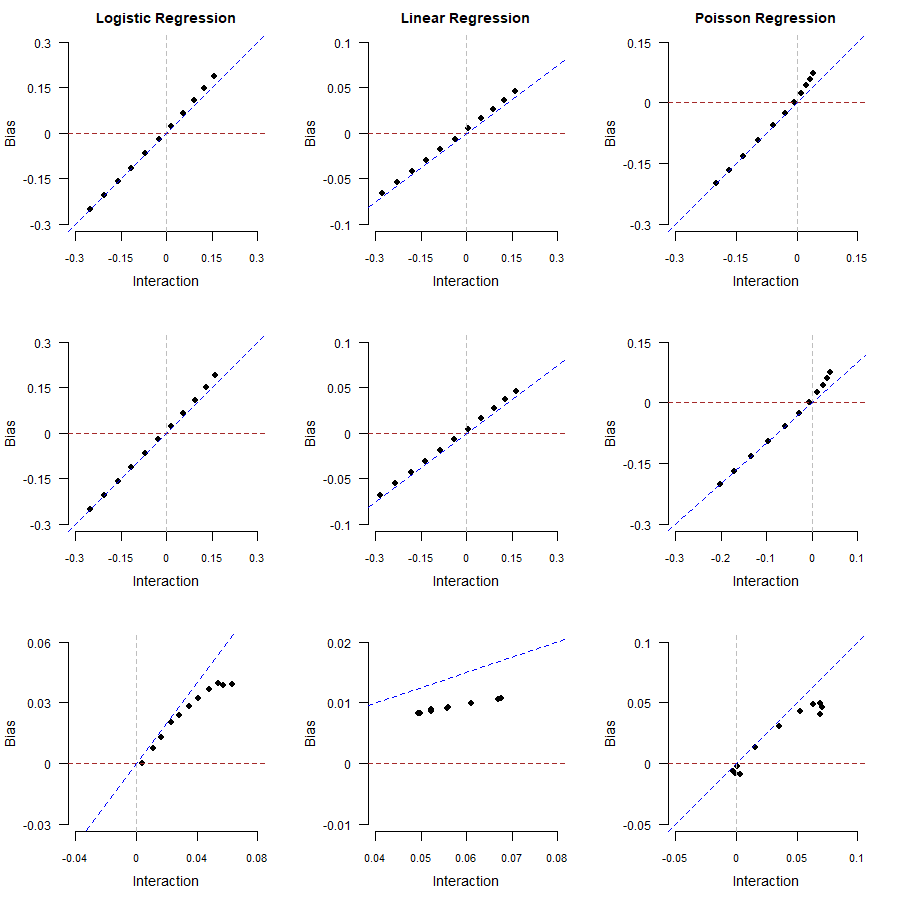}
\caption{Magnitude of collider bias induced in the exposure-outcome regression coefficient by restricting the analysis to selected ($S = 1$) individuals. Exposure data were generated as a continuous variable ($X \sim N(0, 1)$). Outcome data were generated from logistic regression (left column), linear regression (middle column) or Poisson regression (right column) and collider values were generated from logistic regression (model S1, top row), probit regression (model S2, middle row) or the ``double threshold" model (model S3, bottom row). The bias is plotted against the estimated values of the exposure-outcome interaction parameter $\delta_3^{S0}$ in a (misspecified) log-additive model for $S$. A gray vertical line represents no interaction ($\hat{\delta}_3^{S0} = 0$).} \label{SupplAsyPlot4l}
\end{figure}

\begin{figure}[!t]
\centering
\includegraphics[scale = 0.65]{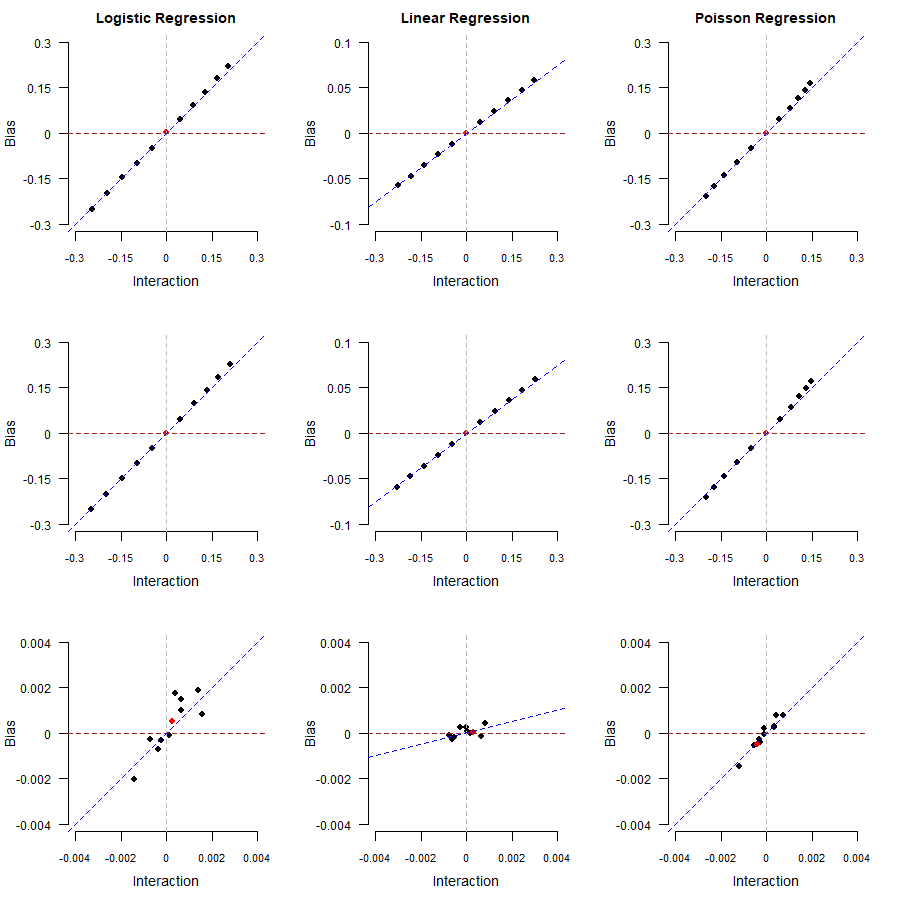}
\caption{Magnitude of collider bias induced in the exposure-outcome regression coefficient by restricting the analysis to selected ($S = 1$) individuals. Exposure data were generated as a continuous variable ($X \sim N(0, 1)$). Outcome data were generated from logistic regression (left column), linear regression (middle column) or Poisson regression (right column) and collider values were generated from logistic regression (model S1, top row), probit regression (model S2, middle row) or the ``double threshold" model (model S3, bottom row). The parameters $\beta_1$ and $\delta_2$ were set to zero. The bias is plotted against the estimated values of the exposure-outcome interaction parameter $\delta_3^{S0}$ in a (misspecified) log-additive model for $S$. Red points represent runs with $\delta_3^{Sk} = 0$, $k = 1, 2, 3$. A grey vertical line represents no interaction ($\hat{\delta}_3^{S0} = 0$).} \label{SupplAsyPlot5z}
\end{figure}

In Figure~\ref{SupplAsyPlot4l}, we have plotted the bias against estimates $\hat{\delta}_3^{S0}$ of the $X-Y$ interaction in the log-additive model (S0). The blue dotted line represents the linear relationship that one would expect to observe based on our theory if the log-additive model was correctly specified. Under misspecification, the relationship between collider bias and $\delta_3^{S0}$ estimates becomes non-linear; this is more pronounced for the ``double-threshold" model but still the case for logistic and probit regression. The deviation from linearity is relatively small for some combinations of outcome and collider models, so the interaction estimate may still carry some value as an informal assessment of the magnitude of collider bias. Nevertheless, the differences between Figure~\ref{SupplAsyPlot4l} and Figure 3 of the main paper suggest that, when the collider is not truly distributed according to the log-additive model, the type of the exposure variable matters in determining the bias.

As a consequence of these results, consider a study where the collider $S$ is distributed according to a logistic regression model, and is affected only by the exposure and not by the outcome:
\begin{equation}
	\text{logit} \mathbb{P} (S = 1 | X, Y) = d_0 + d_1 X \label{s.logit.null}
\end{equation}
The study will not suffer from collider bias, since the absence of a direct $Y-S$ effect means that $S$ s not a collider of the exposure and outcome. However, if one was able to fit the log-additive model \eqref{logadd.s} for $S$, the value of the interaction estimate $\hat{\delta}_3$ in that model would be non-zero due to model misspecification. In this case, using heuristic arguments about the plausibility of exposure-collider and outcome-collider associations in the context of the study may help to assess the presence of collider bias before attempting to quantify it.

In addition, under model \eqref{s.logit.null}, if the (unconditional) exposure-outcome association of interest is also null, this will drive the interaction estimate in the log-additive model towards zero. This is illustrated in Figure~ref{SupplAsyPlot5l} where we have repeated the asymptotic experiment of Figure~\ref{SupplAsyPlot4l} but set both $\delta_2^{Sk}$ and $\beta_1^{Yj}$ to zero($k = 1, 2, 3$, $j = 1, 2, 3$). The bias is plotted against estimated interactions in the log-additive model (S0). In general, these plots are also subject to bias, as in Figure~\ref{SupplAsyPlot4l}. However, in runs where the interaction parameter $\delta_3^{Sk}$ was also set to zero (denoted in red), collider bias does not occur and the interaction estimates in model (S0) were also close to zero. This was the case for all outcome and all collider models. The pattern of bias may seem less clear for the ``double-threshold" model (S3), but this is due to the scale of the axes: there was very little bias in that model regardless of the value of the interaction $\delta_3^{S3}$.

\section{Missingness Rates for Maternal Traits in ALSPAC}

Here, we report the missingness rates for each of the six maternal traits in ALSPAC. The dataset we worked with contained information on $15645$ foetuses in ALSPAC. Of those foetuses, $607$ had missing child sex data, representing miscarriages and early terminations of pregnancy. These were excluded from our analysis, leaving us with $15038$ children. In each of the six univariate analyses, we also excluded families with missing values for the corresponding maternal trait. Missingness rates ranged from $6.1\%$ to $22.8\%$ in the full-ALSPAC sample and were slightly smaller in the TF4 and CCU subsamples. These are summarized in Table~\ref{alspac.missing}.

\begin{table}[!t]
\centering
\begin{tabular}{c|cc|cc|cc}
  \multirow{2}{*}{Mat Trait} & \multicolumn{2}{c}{All ALSPAC} & \multicolumn{2}{c}{TF4} & \multicolumn{2}{c}{CCU} \\
  & Prop & $N$ & Prop & $N$ & Prop & $N$  \\ 
  \hline
  Age 			&  $6.5\%$ &   978 &  $6.4\%$ & 335 &  $4.4\%$ & 191 \\
  Education 	& $17.1\%$ &  2577 &  $9.8\%$ & 510 &  $6.9\%$ & 301 \\
  BMI 			& $22.9\%$ &  3437 & $15.3\%$ & 796 & $12.9\%$ & 559 \\
  Depression 	& $18.0\%$ &  2704 & $10.6\%$ & 554 &  $8.1\%$ & 350 \\
  Smoking 		& $11.8\%$ &  1774 &  $8.4\%$ & 436 &  $5.9\%$ & 255 \\
  Gest Age 		&  $6.1\%$ &   912 &  $6.4\%$ & 335 &  $4.4\%$ & 191 \\
  \hline
  Total 		&   -----  & 15038 &  -----  & 5208 &  -----  & 4333 \\
  \hline
\end{tabular}
\caption{Missingness rates for each of the six maternal traits and sample sizes used in each of the six logistic regression analyses to estimate the association of child sex with the corresponding maternal trait, in the full ALSPAC sample as well as in the two subsamples (TF4 and CCU). \label{alspac.missing}}
\end{table}

\section{Joint Analysis for the Real-data Application}

\begin{center}
\begin{table*}[!t]%
\centering
\scriptsize
\begin{tabular}{c|ccc|ccc|ccc}
  \multirow{2}{*}{Mat Trait} & \multicolumn{3}{c}{All ALSPAC} & \multicolumn{3}{c}{TF4} & \multicolumn{3}{c}{CCU} \\
  & $\hat{\beta}_1$ & s.e. $(\hat{\beta}_1)$ & P-value & $\hat{\beta}_1$ & s.e. $(\hat{\beta}_1)$ & P-value & $\hat{\beta}_1$ & s.e. $(\hat{\beta}_1)$ & P-value \\ 
  \hline
  Age 			&  0.013 & 0.004 & 0.002 &  0.018 & 0.007 & 0.011 &  0.014 & 0.008 & 0.083 \\ 
  Education 	& -0.019 & 0.016 & 0.257 &  0.060 & 0.028 & 0.031 &  0.110 & 0.031 & $3.2 \times 10^{-4}$ \\ 
  BMI 			&  0.002 & 0.005 & 0.692 &  0.014 & 0.008 & 0.911 &  0.012 & 0.010 & 0.221 \\ 
  Depression 	& -0.075 & 0.059 & 0.203 & -0.082 & 0.105 & 0.433 & -0.096 & 0.125 & 0.444 \\ 
  Smoking 		&  0.039 & 0.038 & 0.306 &  0.008 & 0.065 & 0.896 & -0.095 & 0.075 & 0.202 \\ 
  Gest Age 		& -0.054 & 0.011 & $3.6 \times 10^{-7}$ & -0.074 & 0.018 & $2.5 \times 10^{-5}$ & -0.079 & 0.019 & $4.6 \times 10^{-5}$ \\ 
  \hline
\end{tabular}
\caption{Associations between each of six maternal traits and child sex in ALSPAC, obtained either from all ALSPAC participants, or from those who attended the TF4 visit, or from those who returned the CCU questionnaire.\label{Suppl.alspac.all}}
\end{table*}
\end{center}

Finally, we report results of our real-data application when the associations of the six maternal traits with child sex are estimated in a single joint model. Our theory about the connection of collider bias with interactions in log-additive models was presented in terms of models with a single exposure variable, hence the choice to model each maternal trait separately in the real-data application in the main part of our manuscript. However, the theory extends straightforwardly to models with multiple explanatory variables, and this is illustrated here.

Again, our objective was to explore the bias induced in associations between six maternal traits and child sex, when the analysis was restricted to the subsamples of individuals who attended the TF4 clinic visit or those who completed the CCU questionnaire, compared to the overall ALSPAC sample. To do so, we fitted a single logistic regression model for child sex with the six maternal traits as covariates; this was fitted both in the full ALSPAC sample and in the two subsamples. Parameter estimates between the three fits were compared to assess the magnitude of bias that would be induced if the analysis was restricted to the TF4/CCU subsample. 

We limited our analysis to families with fully reported maternal data, and also excluded miscarriages and early terminations of pregnancy. This meant that our full-ALSPAC analysis was conducted on a sample of $10689$ pregnancies ($68.3\%$ of all ALSPAC participants). Of those, $4251$ attended the TF4 visit and $3649$ completed the CCU questionnaire.

Table~\ref{Suppl.alspac.all} contains the results of this analysis. We report parameter estimates, standard errors and p-values of association between each maternal trait and child sex, obtained either from all ALSPAC participants or from those who attended TF4 and CCU. Similar to the marginal analyses in the main part of our paper, age at delivery and gestational age were associated with child sex (though for age at delivery, the association in the CCU sample was weaker). Education showed no evidence of association with child sex in the full ALSPAC sample but did so in the two subsamples, indicating potential collider bias. Unlike the marginal analyses, we did not observe an association of maternal smoking before pregnancy with child sex in any of the three logistic regression fits. Finally, maternal BMI and depression did not associate with child sex.

\begin{center}
\begin{table*}[!t]%
\centering
\scriptsize
  \begin{tabular}{c|cccc|cccc}
  \multirow{2}{*}{Mat Trait} & \multicolumn{4}{c}{TF4} & \multicolumn{4}{c}{CCU} \\
  & Estimate & Bias in $\hat{\beta}_1$ & Std Error & P-value & Estimate & Bias in $\hat{\beta}_1$ & Std Error & P-value \\
  \hline  
Child Sex       	&  0.014  &   ---  & 0.801 & 0.986	&  0.419  &   ---  & 0.895 & 0.639 \\
Age         		&  0.023  &   ---  & 0.005 & $5.0 \times 10^{-7}$	&  0.027  &   ---  & 0.005 & $1.8 \times 10^{-8}$ \\
Education   		&  0.128  &   ---  & 0.018 & $6.8 \times 10^{-13}$	&  0.128  &   ---  & 0.019 & $4.3 \times 10^{-12}$ \\
BMI        			& -0.006  &   ---  & 0.006 & 0.310	& -0.009  &   ---  & 0.006 & 0.113  \\
Depression  		&  0.136  &   ---  & 0.073 & 0.062	&  0.238  &   ---  & 0.080 & 0.003 	\\
Smoking    			& -0.182  &   ---  & 0.046 & $6.5 \times 10^{-5}$	& -0.141  &   ---  & 0.047 & 0.003 \\
Gest Age      		&  0.011  &   ---  & 0.012 & 0.363	&  0.026  &   ---  & 0.013 & 0.042 \\ 
Age x Sex           &  0.002  &  0.005 & 0.007 & 0.721	& -0.003  &  0.001 & 0.008 & 0.716 \\
Education x Sex     &  0.081  &  0.079 & 0.027 & 0.003	&  0.131  &  0.129 & 0.030 & $1.4 \times 10^{-5}$ 	 \\
BMI x Sex           &  0.011  &  0.012 & 0.008 & 0.190	&  0.008  &  0.010 & 0.009 & 0.374 \\
Depression x Sex    & -0.030  & -0.007 & 0.106 & 0.778	& -0.058  & -0.021 & 0.123 & 0.634 \\
Smoking x Sex       & -0.034  & -0.031 & 0.070 & 0.632	& -0.140  & -0.134 & 0.078 & 0.073 \\
Gest Age x Sex     	& -0.020  & -0.020 & 0.017 & 0.248	& -0.028  & -0.025 & 0.019 & 0.140 \\
  \hline
\end{tabular}
\caption{Parameter estimates, standard errors and p-values for a log-additive model of TF4 or CCU participation in terms of child sex, maternal traits and interactions between child sex and maternal traits. The observed bias in $\hat{\beta}_1$ estimates from Table~\ref{Suppl.alspac.all} is also reported for comparison. \label{Suppl.alspac.logadd}}
\end{table*}
\end{center}

To explore whether these associations were due to collider bias, we then fitted a log-additive model for TF4/CCU participation in terms of child sex, the six maternal traits and interactions between child sex and the maternal traits. We did not include interactions between the maternal traits, as our theory suggests that such interactions play no role in determining collider bias. The interaction parameter estimates from the log-additive model were then compared to the observed bias in ALSPAC. Results are reported in Table~\ref{Suppl.alspac.logadd}

All maternal traits apart from BMI were associated with CCU participation, and three maternal traits (education, age at delivery and smoking) were associated with TF4 participation. However only for maternal education was there evidence of an interaction with child sex in the log-additive model for participation. This was in line with our previous analysis, in which maternal education was the only trait that associated with child sex among CCU or TF4 participants but not in the complete ALSPAC sample. In addition, for all maternal traits, interaction parameter estimates were again similar to the observed bias in the ALSPAC analyses, computed from Table~\ref{Suppl.alspac.all} as the difference between regression parameter estimates in TF4/CCU samples and in the full ALSPAC sample. A notable difference between bias and interaction parameter estimates was observed for the depression-child sex interaction, but the standard errors for the corresponding parameter estimate were larger than for other traits and the difference can likely be attributed to random variation.

\begin{center}
\begin{table*}[!t]%
\centering
\footnotesize
\begin{tabular}{c|ccc|ccc}
  \multirow{2}{*}{Mat Trait} & \multicolumn{3}{c}{TF4} & \multicolumn{3}{c}{CCU} \\
  & Estimate & Std Error & P-value & Estimate & Std Error & P-value \\ 
  \hline
Child Sex  	& -0.510  & 0.041 & $3.8 \times 10^{-35}$	& -0.747  & 0.043 & $9.7 \times 10^{-68}$ \\
Age         &  0.043  & 0.005 & $4.2 \times 10^{-21}$	&  0.043  & 0.005 & $4.0 \times 10^{-19}$ \\
Education   &  0.283  & 0.018 & $1.4 \times 10^{-57}$ 	&  0.288  & 0.019 & $1.4 \times 10^{-54}$ \\
BMI         & -0.002  & 0.006 & 0.702 					& -0.010  & 0.006 & 0.083 \\
Depression  &  0.194  & 0.066 & 0.004					&  0.318  & 0.073 & $1.2 \times 10^{-5}$ \\
Smoking     & -0.311  & 0.043 & $6.1 \times 10^{-13}$	& -0.286  & 0.045 & $2.9 \times 10^{-10}$ \\
Gest Age    &  0.002  & 0.011 & 0.829 					&  0.022  & 0.012 & 0.067 \\
  \hline
\end{tabular}
\caption{Parameter estimates, standard errors and p-values for a logistic model of TF4 and CCU participation in terms of child sex and maternal traits. \label{Suppl.alspac.logit}}
\end{table*}
\end{center}

As in the main text, we also implemented logistic regression for TF4 and CCU participation using child sex and the six maternal traits as explanatory variables but no interaction terms. The results of fitting the logistic regression model to the ALSPAC data are given in Table~\ref{Suppl.alspac.logit}. We report parameter estimates, standard errors and p-values for child sex and each maternal variable.

Associations in this joint analysis are generally weaker than in the marginal analyses presented in the main part of our paper, but nevertheless, mother's age at pregnancy, education, depression and smoking before pregnancy all associate with participation into the TF4/CCU samples. As with the marginal analyses, this could lead researchers using the logistic model to worry about collider bias affecting all four of these traits. However, as Table~\ref{Suppl.alspac.all} suggests, only for maternal education is there a cause for concern.
